\documentclass{iopart}

\usepackage{graphicx, verbatim,cite,color,array}

\newcommand{\ket}[1]{\left\vert{#1}\right\rangle}

\begin{document}

\title{Simulating chemistry efficiently on fault-tolerant quantum computers}
\author{N. Cody Jones$^{1,*}$,
        James D. Whitfield$^{2,3,4}$,
        Peter L. McMahon$^{1}$,
        Man-Hong Yung$^{2}$,
        Rodney Van Meter$^{5}$,
        Al\'{a}n Aspuru-Guzik$^{2}$,
        and Yoshihisa Yamamoto$^{1,6}$}
\address{$^1$
         Edward L. Ginzton Laboratory,
         Stanford University,
         Stanford, CA 94305-4088, USA}
\address{$^2$
         Harvard University, Department of Chemistry and Chemical Biology, 12 Oxford St., Cambridge, MA 02138, USA}
\address{$^3$
         NEC Laboratories, America, 4 Independence Way, NJ 08540, USA}
\address{$^4$
         Columbia University, Physics Department, 538 W $120^{\mathrm{th}}$ St., New York, NY 10027, USA}
\address{$^5$
        Faculty of Environment and Information Studies, Keio University, Japan}
\address{$^6$
         National Institute of Informatics,
         2-1-2 Hitotsubashi, Chiyoda-ku, Tokyo 101-8430, Japan}

\address{$^{*}$
         Corresponding author:}
         \ead{ncodyjones@gmail.com}

\begin{abstract}
Quantum computers can in principle simulate quantum physics exponentially faster than their classical counterparts, but some technical hurdles remain.  Here we consider methods to make proposed chemical simulation algorithms computationally fast on fault-tolerant quantum computers in the circuit model.  Fault tolerance constrains the choice of available gates, so that arbitrary gates required for a simulation algorithm must be constructed from sequences of fundamental operations.  We examine techniques for constructing arbitrary gates which perform substantially faster than circuits based on the conventional Solovay-Kitaev algorithm [C.M. Dawson and M.A. Nielsen, \emph{Quantum Inf. Comput.}, \textbf{6}:81, 2006].  For a given approximation error $\epsilon$, arbitrary single-qubit gates can be produced fault-tolerantly and using a limited set of gates in time which is $O(\log \epsilon)$ or $O(\log \log \epsilon)$; with sufficient parallel preparation of ancillas, constant average depth is possible using a method we call programmable ancilla rotations.  Moreover, we construct and analyze efficient implementations of first- and second-quantized simulation algorithms using the fault-tolerant arbitrary gates and other techniques, such as implementing various subroutines in constant time.  A specific example we analyze is the ground-state energy calculation for Lithium hydride.
\end{abstract}



\section{Introduction}
Simulating quantum physics is arguably one of the most important applications of a quantum computer---a problem whose solution is both intractable for classical computers and valuable to science~\cite{Feynman82}.  The objective of quantum simulation is to model natural physical systems with Hamiltonians that permit a compact representation~\cite{Lloyd1996,Zalka1998b}.  In this investigation, we narrow our focus to quantum chemistry problems such as calculating the eigenvalues of a molecular Hamiltonian \cite{Aspuru05,wang2008quantum,Veis2010,Whitfield2011}.  We aim to demonstrate constructively how quantum computers can simulate chemistry with an efficient use of resources.  By doing so, we indicate how close the field of quantum information processing is to solving novel problems for less computational cost than a classical computer.

Quantum chemistry and band structure calculations account for up to 30\% of the computation time used at supercomputer centers~\cite{NERSC2010}. The most-employed methods include density functional theory and polynomially-tractable approximate quantum chemistry methods~\cite{HeadGordon2008}.  Despite the success of these methods, for example, in simulating the dynamics of a small protein from first principles~\cite{Ufimtsev2011} or in predicting novel materials~\cite{Sokolov2011}, they are still approximate, and much work is carried out in developing more accurate methods.  Quantum simulators offer a fresh approach to quantum chemistry~\cite{Kassal2011} as they are predicted to allow for the exact simulation (within a basis) of a chemical system in polynomial time. A quantum computer of a sufficient size, say 128 logical quantum bits~\cite{Aspuru05,Kassal2008}, would already outperform the best classical computers for \emph{exact} chemical simulation.  This would open the door to high-quality \emph{ab initio} data for parameterizing force fields for molecular dynamics~\cite{Huang12} or understanding complex chemical mechanisms such as soot formation~\cite{Wheeler07}, where a number of different chemical species must be compared. This tends to suggest that computational chemistry would be one of the first novel applications of universal quantum computers.

The motivation behind our study is that in order for computational physics on quantum computers to be useful as a scientific tool, it must have an efficient implementation. Often general algorithmic complexity such as ``polynomial time'' is taken as a by-word for efficient, but we go deeper to show the substantial performance disparities between different polynomial-time algorithms, revealing which ones are significantly more efficient in space and time resources than their peers.  By introducing algorithmic improvements and making quantitative analysis of the resource costs, we show that simulating quantum chemistry is feasible in a practical execution time, such as simulating the ground state energy of Lithium hydride (LiH) in $\sim 5.6$ hours on a hypothetical fault-tolerant quantum computer with an execution time per error-corrected gate of \mbox{1 ms}.  Additionally, from an information theory perspective, it is interesting to see what quantum computational complexity is required to simulate physically-relevant Hamiltonians in general~\cite{Poulin2011}.

Several possible simulators have been proposed and studied~\cite{Buluta2009,Kassal2011,Barreiro2011,Simon2011,Ma2011}, but we focus on fault-tolerant circuit-model quantum simulation in this investigation~\cite{Lloyd1996,Aspuru05,Clark09,Kassal2008,Lanyon2010,Brown2010,Whitfield2011,Lanyon2011}.  The reasons for these constraints are straightforward: quantum computers will probably be sensitive to noise and other hardware errors, thus requiring fault tolerance~\cite{Preskill97}, and fault-tolerant quantum computing has been most successfully applied in the circuit-model.  Fault tolerance requires an overhead of additional work for the quantum computer; error correcting codes and the mechanisms they use to correct errors have been studied previously~\cite{Preskill97,Nielsen00,Devitt2009}.  We focus here on another matter critical to simulation algorithms, which is making arbitrary fault-tolerant gates.  Arbitrary quantum operations, such as a single-qubit rotation of arbitrary angle around the $Z$-axis on the Bloch sphere, are typically constructed using a sequence of primitive error-corrected gates~\cite{Nielsen00,Dawson05,Fowler2011}.  Quantum simulation depends sensitively on the execution time of arbitrary gates of this form, so one of the core contributions of this paper is to demonstrate efficient constructions for such gates, which would allow simulation of more complex systems under a fixed-resource constraint.

\begin{figure}
  \centering
  \includegraphics[width=\textwidth]{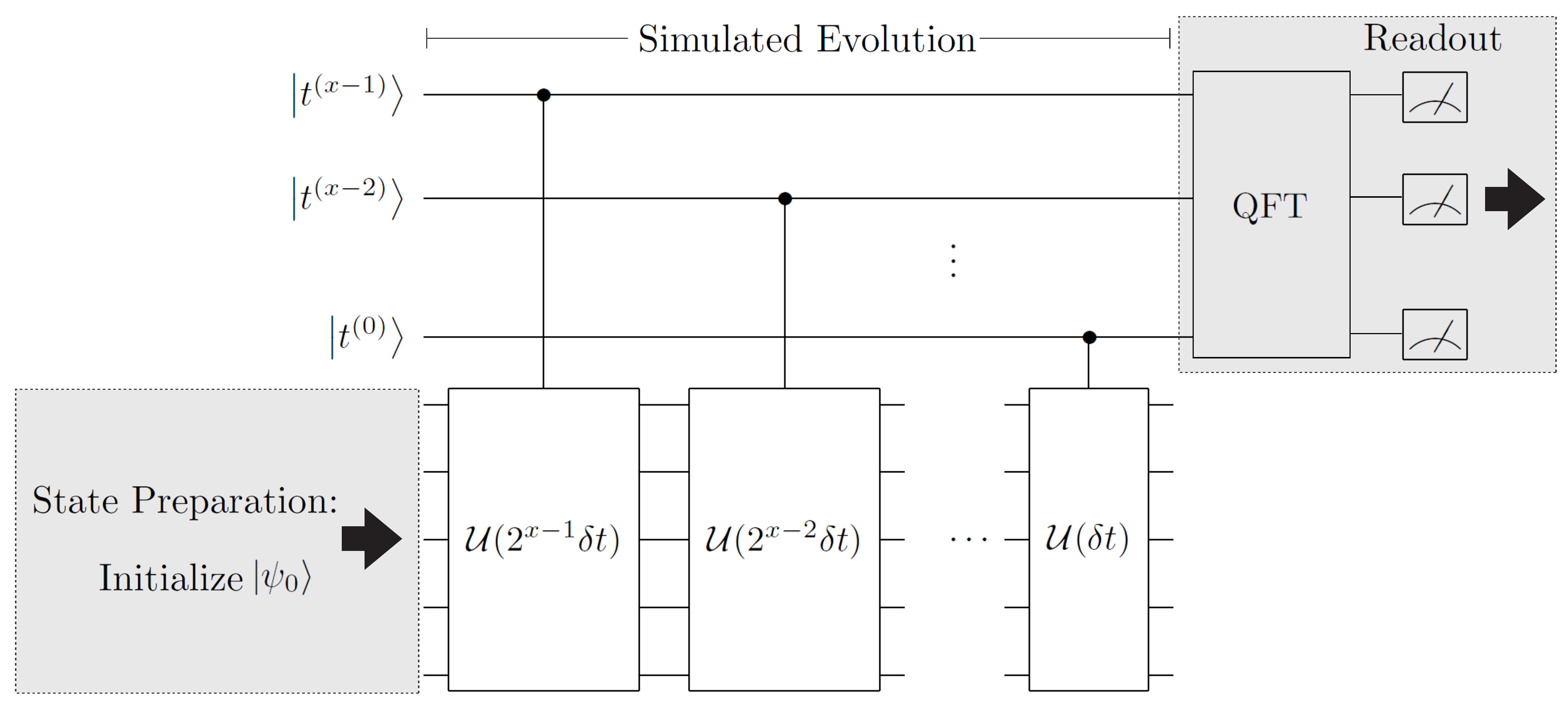}\\
  \caption{Schematic of a digital quantum simulation algorithm for energy eigenvalue estimation~\cite{Lloyd1996,Aspuru05}.  The three main steps are state preparation, simulated evolution, and readout; this investigation focuses on the middle process.  After preparing an initial state $\ket{\psi_0}$, the system is evolved in simulated time by solving the time-dependent Schr\"{o}dinger equation.  Note the system propagators $\mathcal{U}(2^x\delta t)$ are controlled by qubits in a register representing simulated time.  A quantum Fourier transform (QFT) on the time register provides an estimate of an energy eigenvalue.  The accuracy of the simulation depends on suppressing errors in both state preparation and simulated-time evolution, which is why fault tolerance is an important consideration for quantum simulation algorithms.}
  \label{sim_algorithm}
\end{figure}

A digital quantum simulation algorithm consists of three primary steps (figure~\ref{sim_algorithm}): state preparation, simulated time evolution, and measurement readout.  This paper focuses on the second step, evolving the system in simulated time, because this represents the core of the algorithm.  Simulation of time evolution on a quantum computer is a sequence of quantum gates which closely approximates the evolution propagator $\mathcal{U}(t;t+\delta t) = \mathcal{T}\exp\left(-\frac{i}{\hbar}\int_{t}^{t+\delta t}\mathcal{H(\tau)}d\tau\right)$ of a desired Hamiltonian $\mathcal{H}$, where $\mathcal{T}$ is the usual time-ordering operator.  In the case of a time-independent Hamiltonian, we have $\mathcal{U}(\delta t) = \exp\left(-\frac{i}{\hbar} \mathcal{H} \delta t\right)$, as in figure~\ref{sim_algorithm}.  The increment $\delta t$ is a single time step of simulation, and a simulation algorithm often requires many time steps, depending on the desired result (\emph{e.g.} energy eigenvalue).  State preparation and measurement readout are necessary steps which are not discussed here, but details can be found in references~\cite{Abrams1997,Zalka1998b,Abrams1999,Nielsen00,Grover2002,Mohseni2006}.

The quantum simulation problem we analyze is the ground-state energy calculation of LiH from first principles. This was called the ``chemist's workbench'' and is an appropriate continuation of quantum computational applications of chemistry going beyond molecular Hydrogen~\cite{Ben88,Lanyon2010,Du2010,Whitfield2011}.  For some of the selected methods, the quantum circuit is compact enough to be tractable for classical computation, so our chosen problem would not demonstrate the superiority of quantum computation by itself.  Still, this example is useful for two reasons.  First, the LiH simulation preserves the features of more complicated chemical simulations while permitting a simple analysis that illustrates the improved methods we propose.  Second, with quantum computers still in early stages of development, a compact problem such as LiH would be a convenient choice for experimental demonstrations of quantum simulation in the near-term.

This paper provides constructive methods for simulating quantum chemistry efficiently using fault-tolerant quantum circuits.  Section~\ref{phase_gates} describes how to construct quantum circuits for arbitrary phase rotations, which are essential to simulation.  Section~\ref{second_quantized} develops a fault-tolerant simulation algorithm in second-quantized representation using phase rotations from the prior section; analysis of the computing resources required follows. Section~\ref{first_quantized} demonstrates how to construct an efficient chemistry simulation in first-quantized form, and total quantum resources are analyzed.  Section~\ref{comparing_methods} outlines how to determine the optimal simulation parameters for a given set of engineering constraints and performance objectives.  The paper concludes by discussing the prospects for fault-tolerant quantum computers to solve novel simulation problems.

\section{Fault-tolerant phase rotations}
\label{phase_gates}
The algorithms which simulate chemistry on a circuit-model quantum computer require many phase rotations, accurate to high-precision.  A single-qubit rotation gate in general form is
\begin{equation}
\label{phase_rotation}
R_Z(\phi) = e^{i\frac{\phi}{2}} e^{-i\frac{\phi}{2}\sigma_Z} = \left[\begin{array}{cc} 1 & 0 \\ 0 & e^{i\phi} \end{array} \right],
\end{equation}
where $\phi$ is arbitrary and $\sigma_Z$ is the Pauli spin operator; in general, phase rotations are represented by diagonal unitary matrices, as shown on the RHS of Eqn.~(\ref{phase_rotation}).  Additionally, any arbitrary single-qubit gate can be produced using three distinct phase rotations and two Hadamard gates~\cite{Nielsen00}.  Making a quantum computer fault-tolerant constrains the available operations to a finite set of fundamental gates, so the arbitrary rotations needed to simulate Hamiltonian evolution must be constructed from a circuit of these fundamental gates.  Phase rotations are needed at every time step of simulation, so the performance of the simulation algorithm depends on the computational complexity of these arbitrary gate circuits.  In this section, we discuss three different approaches for implementing arbitrary phase gates efficiently: \emph{phase kickback}~\cite{Kitaev1995,Cleve1998,Kitaev2002}, which uses multi-qubit gates acting on an ancilla register; \emph{gate approximation sequences}, such as those generated by the Solovay-Kitaev algorithm~\cite{Nielsen00,Dawson05} or by Fowler's algorithm~\cite{Fowler2011}, which are sequences of single-qubit gates; and \emph{programmable ancilla rotations} (PARs), which compute ancillas in advance using one of the above methods to achieve very low circuit depth in the algorithm.

\subsection{Phase kickback}
\label{pk_section}
Phase kickback~\cite{Kitaev1995,Cleve1998}, also known as the Kitaev-Shen-Vyalyi algorithm~\cite{Kitaev2002}, is an ancilla-based scheme that uses an addition circuit to impart a phase to a quantum register.  Phase kickback relies on a resource state $\left|\gamma^{(k)}\right\rangle$ which can be defined by the inverse quantum Fourier transform (QFT)~\cite{Nielsen00,Cleve2000,Weinstein2001}:
\begin{equation}
\label{QFT}
|\gamma^{(k)}\rangle = {U_{\mathrm{QFT}}}^{\dag}\left|k\right\rangle = \frac{1}{\sqrt{N}}\sum_{y=0}^{N-1}e^{-2\pi i ky/N}\left|y\right\rangle.
\end{equation}
The register $\ket{k}$ contains $n$ qubits prepared in the binary representation of $k$, an odd integer.  The state $\ket{\gamma^{(k)}}$ is a uniform-weighted superposition state containing the ring of integers from $0$ to $N-1$, where $N = 2^n$, and each computational basis state has a relative phase proportional to the equivalent binary value of that basis state.  This ancilla register must be produced fault-tolerantly.  Ref.~\cite{Kitaev2002} provides a method to prepare $\ket{\gamma^{(k)}}$ using phase estimation such that $k$ is a random odd integer; hence our analysis does not assume a value for $k$.  If necessary, \ref{transform_gamma} provides a technique to convert any $\ket{\gamma^{(k)}}$ into $\ket{\gamma^{(1)}}$.  The circuit complexity for creating $\ket{\gamma^{(k)}}$ is small, requiring perhaps a few thousand gates, so the cost of this initialization step is negligible compared to quantum algorithms we analyze later.

One could also view the $\left|\gamma^{(k)}\right\rangle$ state as a discretely-sampled plane wave with wavenumber $k$.  Consider then that $\left|\gamma^{(k)}\right\rangle$ is an eigenstate of the unitary operation $U_{\oplus u}\ket{m} = \ket{m + u \; (\mathrm{mod} \; N)}$ for modular addition, so that
\begin{equation}
U_{\oplus u}|\gamma^{(k)}\rangle = \frac{1}{\sqrt{N}}\sum_{y=0}^{N-1}e^{2\pi i k(u-y)/N}\left|y\right\rangle = e^{2\pi i ku/N}|\gamma^{(k)}\rangle,
\label{PK_definition}
\end{equation}
where $\oplus$ denotes addition modulo $N$ and $u$ is an integer.  Moreover, the eigenvalue of modular addition on $\left|\gamma^{(k)}\right\rangle$ is a phase factor proportional to the number $u$ added.  Note that the addition operation $U_{\oplus u}$ is readily implemented with a fault-tolerant quantum circuit~\cite{Vedral1996,Draper2000,VanMeter05,Cuccaro2004,Draper2006}.  To determine the value of $u$ in the addition circuit which approximates a phase rotation $R_Z(\phi)$, one solves the modular equation
\begin{equation}
\label{mod_equation}
ku \equiv \left\lfloor N \frac{\phi}{2\pi} \right\rceil \; (\mathrm{mod} \; N),
\end{equation}
which always has a solution since $k$ is odd and $N$ is a power of $2$.  The operation $\lfloor x \rceil$ denotes rounding any real $x$ to the nearest integer; any arbitrary rule for half-integer values suffices here.  By proper selection of $u$, one can approximate any phase rotation to within a precision of $|\Delta\phi| \le \frac{2\pi}{2^{n+1}}$ radians, where $\Delta\phi = \left[\phi - \frac{2\pi}{N}ku \; (\mathrm{mod} \; 2\pi)\right]$.  We can now understand how the method received its name: since $\left|\gamma^{(k)}\right\rangle$ is an eigenstate of addition, when an integer $u$ is added (using an addition circuit) to this register, a phase is ``kicked back.''  This method is quite versatile, as several different types of phase gates are developed using phase kickback in this work.

\begin{figure}
  \centering
  %
  %
  %
  %
  %
  \includegraphics[width=9cm]{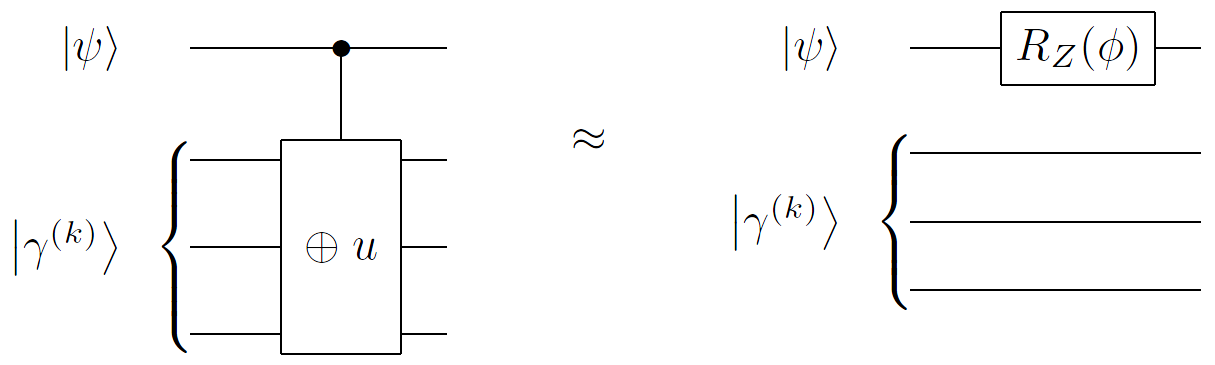}\\
  \caption{Controlled addition of the quantity $u$ determined by Eqn.~(\ref{mod_equation}) is approximately equivalent to an arbitrary phase rotation $R_Z(\phi)$, but the former uses only fault-tolerant gate primitives and ancillas.  The operation $\oplus$ denotes unitary addition modulo $2^n$, where $n$ is the number of qubits in the $\ket{\gamma^{(k)}}$ register; for illustration, $n = 3$ in the circuits above.}
  \protect\label{single_qubit_PK}
\end{figure}

Single-qubit phase rotations using phase kickback are constructed with a controlled addition circuit, as shown in figure~\ref{single_qubit_PK}.  Intuitively, a phase is kicked back to the control qubit if it is in the $\ket{1}$ state, which is equivalent to the phase rotation in Eqn.~(\ref{phase_rotation}).  The accuracy of the phase gate and the quantum resources required depend on the number of bits in the ancilla state $\left|\gamma^{(k)}\right\rangle$.  After solving Eqn.~(\ref{mod_equation}), the integer $u$ is added to $\left|\gamma^{(k)}\right\rangle$ using a quantum adder controlled by the qubit which is the target of the phase rotation.  There are various implementations of quantum adder circuits which have tradeoffs in performance between circuit depth and circuit size~\cite{Vedral1996,Draper2000,VanMeter05,Cuccaro2004,Draper2006}.  Since $\left|\gamma^{(k)}\right\rangle$ is not altered by phase kickback, the number of such registers required for a quantum algorithm is equal to the maximum number of phase rotations which are computed in parallel at any point in the algorithm.

\subsection{Gate approximation sequences}
\label{app_sequences}
A gate approximation sequence uses a stream of fault-tolerant single-qubit gates to approximate an arbitrary phase rotation, such as that in Eqn.~(\ref{phase_rotation}).  For context, a common set of fault-tolerant gates is listed in Table~\ref{gate_set} below.  Such sequences must be calculated using a classical algorithm, and at least two options exist.  The Solovay-Kitaev algorithm~\cite{Nielsen00,Dawson05} is perhaps the best known method for generating arbitrary quantum operations, so it will serve as a benchmark in our analysis.  A subsequently-derived alternative, Fowler's algorithm~\cite{Fowler2011}, offers shorter gate sequences for a given approximation accuracy, with some notable drawbacks in classical algorithmic complexity.

The efficiency of a gate approximation sequence is determined by the accuracy of approximation (\emph{i.e.} how close the composite sequence is to the desired gate) as a function of resource costs.  Both the Solovay-Kitaev and Fowler algorithms produce better approximations if one can afford more quantum gates; however, quantum resources are expensive, so we must implement finite-length sequences which produce a sufficiently good approximation.  We adopt the distance measure in Ref.~\cite{Fowler2011} to determine approximation accuracy:
\begin{equation}
\textrm{dist}_d(U,V) = \sqrt{\frac{d - \left| \mathrm{tr}(U^{\dag}V)\right|}{d}},
\label{distance_metric}
\end{equation}
where $d$ is the dimensionality of $U$ and $V$ (\emph{e.g.} $d = 2$ for a single-qubit rotation).  At the end of this section, we provide a quantitative analysis of resource costs to produce phase rotations.  What is sufficient for the moment is to know that, if we denote the approximation error as $\epsilon = \mathrm{dist}_2(U,U_{\mathrm{approx}})$, the corresponding approximating sequence $U_{\mathrm{approx}}$ has asymptotic length $O(\mathrm{poly}(\log \epsilon))$, a result known as the Solovay-Kitaev theorem~\cite{Nielsen00}.

\subsection{Programmable ancilla rotation}
\label{par_section}
We introduce a third method for producing phase rotations, the programmable ancilla rotation (PAR), which pre-computes ancillas before they are needed.  Shifting the computing effort to a different point in the quantum circuit (assuming parallel computation) allows this method to achieve \emph{constant average depth} in the algorithm for any desired accuracy of rotation, which can be as small as 4 quantum gates. The pre-calculated ancillas still require quantum circuits of similar complexity to the previously discussed methods, so this approach is best-suited to a quantum computer with many excess qubits for parallel computing.

\begin{figure}
  \centering
  \includegraphics[width=9cm]{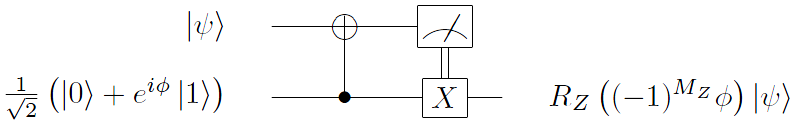}\\
  \caption{Probabilistic rotation using an ancilla qubit.  The measurement is in the computational (\texttt{Z}) basis.  The circuit enacts either $R_Z(\phi)$ or $R_Z(-\phi)$ with equal probability.  The \texttt{X} gate is classically conditioned on the measurement result.}
  \protect\label{ancilla_rotation}
\end{figure}

The PAR is based on a simple circuit which uses a single-qubit ancilla to make a phase rotation, which is a ``teleportation gate''~\cite{Gottesman1999,Zhou2000}, as shown in figure~\ref{ancilla_rotation}.  This circuit is probabilistic, so there is a 50\% probability of enacting $R_Z(-\phi)$ instead of $R_Z(\phi)$; in such an event, we attempt the circuit again with angle $2 \phi$, then $4 \phi$ if necessary, \emph{etc}.  This proceeds until the first observation of a positive angle rotation, in which case we have enacted a rotation $\phi_{\mathrm{total}} = 2^m \phi - \sum_{x=1}^{m-1}2^x\phi = \phi$.

\begin{figure}
  \centering
  \includegraphics[width=\textwidth]{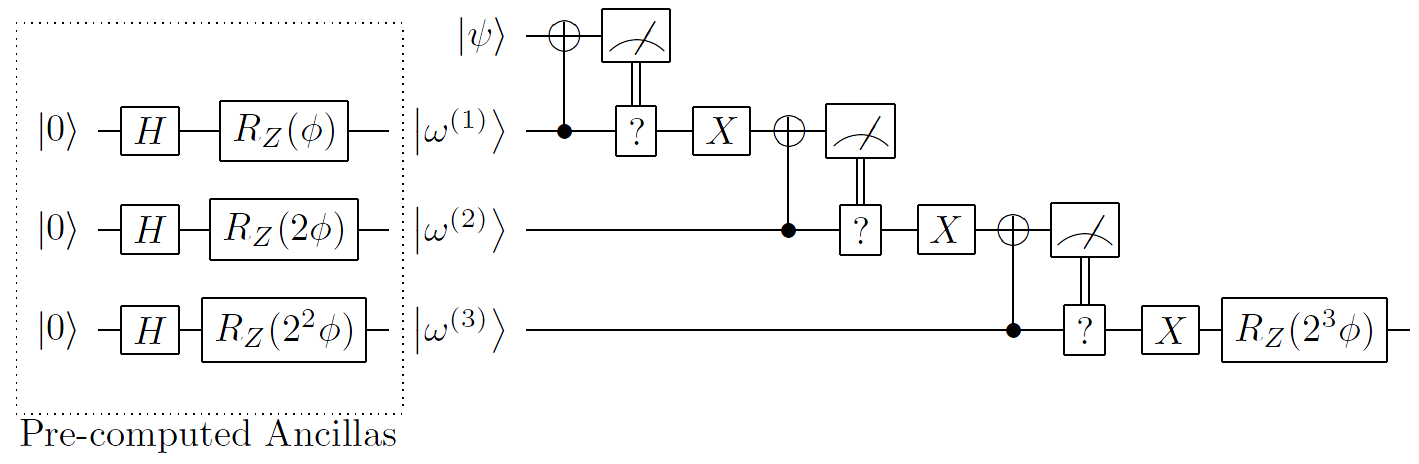}\\
  \caption{Programmable ancilla rotation (PAR) circuit.  The bulk of the computing effort is shifted to an earlier part of the circuit, when the ancillas are produced.  The programmed ancillas are used in multiple rounds of the circuit in figure~\ref{ancilla_rotation}, each of which succeeds with 50\% probability.  The cascading circuit above terminates after the first success, as denoted by the ``?'' decision gates.  The average number of rounds required is 2, so by pre-computing the ancillas, this method contributes very few additional gates to an algorithm's circuit depth.}
  \protect\label{PAR_circuit}
\end{figure}

The circuit for the PAR is shown in figure~\ref{PAR_circuit}.  The programmed ancillas $\ket{\omega^{(1)}} = \frac{1}{\sqrt{2}}\left(\ket{0}+e^{i\phi}\ket{1}\right)$, $\ket{\omega^{(2)}} = \frac{1}{\sqrt{2}}\left(\ket{0}+e^{i(2\phi)}\ket{1}\right)$, \emph{etc}. are pre-computed using one of the methods above for a phase rotation.  A very similar method was shown in Ref.~\cite{Isailovic08}, but we generalize here from $\phi = \frac{\pi}{2^k}$ to arbitrary rotation angles.  In practice, phase kickback may be preferable for producing the pre-computed ancillas since reusing the same $\left|\gamma^{(k)}\right\rangle$ ancilla does not introduce additional errors into the circuit.  The cascading series of probabilistic rotations continues until the desired rotation is produced or the programmed ancillas are exhausted.  For practical reasons, one may only calculate a finite number of the PAR ancillas, and if all such rotations fail, then a deterministic rotation using phase kickback or a gate approximation sequence is applied.  The probability of having to resort to this backstop is suppressed exponentially with the number of PAR ancillas pre-computed.

The average number of rounds of the circuit in figure~\ref{PAR_circuit} before a successful rotation is simply given by $\sum_{m = 1}^{\infty} \frac{m}{2^m} = 2$.  The \texttt{X} gate in each round can be performed with a Pauli frame~\cite{Knill05,DiVincenzo07,Jones2011}, so counting measurement as a gate, the number of gates per round is 2, and the average number of gates per PAR is 4.  With a finite number of pre-computed ancillas $M$, there is a probability $2^{-M}$ of having to implement the considerably more expensive (in circuit depth) deterministic rotation.  Nevertheless, if the computer supports the ability to calculate the programmed ancillas in advance, the PAR produces phase rotations that are orders of magnitude faster than other available methods, which also leads to faster simulation algorithms.

\subsection{Analysis of a single-qubit phase rotation}
\label{single_qubit_rotations}
We begin our quantitative analysis by examining fault-tolerant single-qubit phase rotations.  We construct rotations using phase kickback, the Solovay-Kitaev algorithm, Fowler's algorithm, and PARs.  In each case, we determine the depth of the quantum circuit and the types of fault-tolerant gates required.  The techniques developed here will be used in the more complicated phase rotations for the simulation algorithms in Sections~\ref{second_quantized} and~\ref{first_quantized}.


\begin{table}
  \centering
      \begin{tabular}{| m{2cm} | m{3cm} | m{6cm} |}
      \hline
      \textbf{Symbol} & \textbf{Name} & \textbf{Matrix Representation}\\ \hline
      \texttt{X}, \texttt{Y}, \texttt{Z} & Pauli gates & $\left[\begin{array}{cc} 0 & 1 \\ 1 & 0 \end{array} \right]$, $\left[\begin{array}{cc} 0 & -i \\ i & 0 \end{array} \right]$, $\left[\begin{array}{cc} 1 & 0 \\ 0 & -1 \end{array} \right]^{^{^{\;}}}_{_{\;}}$ \\ \hline
      \texttt{H} & Hadamard & $\frac{1}{\sqrt{2}}\left[\begin{array}{cc} 1 & 1 \\ 1 & -1 \end{array} \right]^{^{^{\;}}}_{_{\;}}$ \\ \hline
      \texttt{S} & $\pi/4$ phase gate & $\left[\begin{array}{cc} 1 & 0 \\ 0 & i \end{array} \right]^{^{^{\;}}}_{_{\;}}$ \\ \hline
      \texttt{T} & $\pi/8$ phase gate & $\left[\begin{array}{cc} 1 & 0 \\ 0 & e^{i\pi/4} \end{array} \right]^{^{^{\;}}}_{_{\;}}$ \\ \hline
      \texttt{CNOT} & Controlled-NOT \newline (two-qubit gate) & $\left[\begin{array}{cccc} 1 & 0 & 0 & 0 \\ 0 & 1 & 0 & 0 \\ 0 & 0 & 0 & 1 \\ 0 & 0 & 1 & 0 \end{array} \right]^{^{^{\;}}}_{_{\;}}$ \\ \hline
      \end{tabular}
  \caption{Universal set of fault-tolerant gates in this investigation.}
  \label{gate_set}
\end{table}


To assess the performance of quantum circuits, let us assume the following simplified quantum computing model.  The hypothetical system uses fault-tolerant quantum error correction, so we presume the quantum gates are ideal.  The quantum computer only has access to a limited set of ``fundamental'' gates, which are summarized in Table~\ref{gate_set}; this set of gates is typical for a fault-tolerant quantum computer~\cite{Nielsen00,Isailovic08,Fowler09,Jones2011}.  We allow full parallelism so that gates can be applied to all qubits simultaneously, as long as the two-qubit (\texttt{CNOT}) gates do not overlap.  Because the fundamental gate set has a finite number of members, phase kickback or gate approximation sequences are required to produce approximations to arbitrary gates.  We should note that each logical gate with error correction will require many more physical operations to implement~\cite{Preskill97,Isailovic08,Jones2011}, but we purposefully avoid these details so that our present analysis is independent of hardware and error correction models.

\begin{figure}
  \centering
  \includegraphics[width=\textwidth]{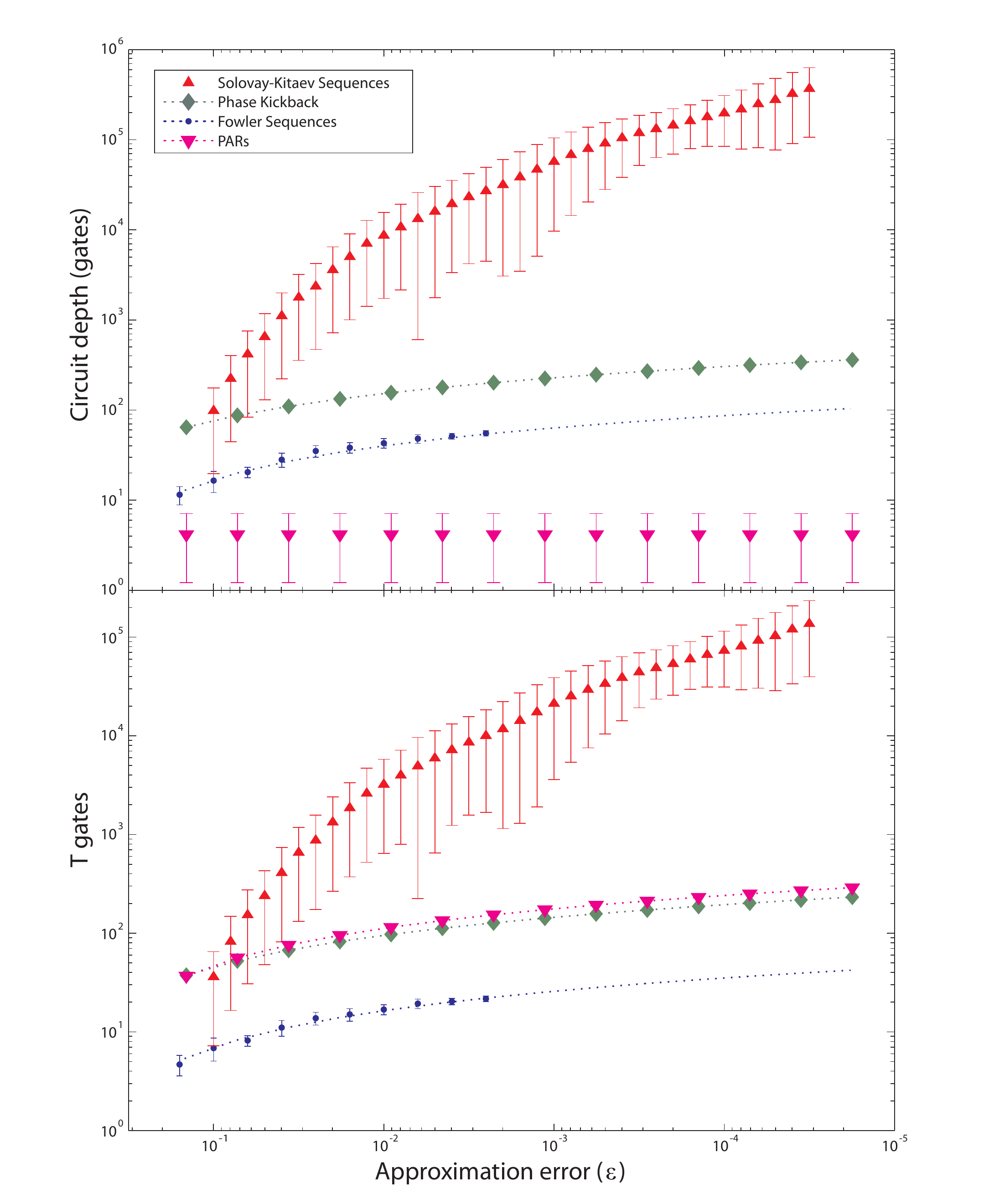}\\
  \caption{Color.  Quantum computing resources required to produce a fault-tolerant single-qubit phase rotation to accuracy $\epsilon = \mathrm{dist}_2\left(R_Z(\phi),U_{\mathrm{approx}}\right)$ using various methods.  \textbf{(top)} Circuit depth for single-qubit rotations.  \textbf{(bottom)} Number of \texttt{T} gates required for each rotation.  There is variation in the resources required for Solovay-Kitaev sequences, Fowler sequences, and PARs; each point is the mean number of gates required, and where applicable, the bars show plus/minus one standard deviation.  The Solovay-Kitaev data is averaged over 9534 random angles ($\phi$), and the Fowler data is averaged over 98 random angles per point.  Fowler sequences are numerically intensive to calculate, so curves fit to the data are shown for $\epsilon \le 10^{-3}$: depth $= -24.9 \log_{10} \epsilon - 7.64$ and \texttt{T} gates $= -9.75 \log_{10} \epsilon - 2.81$.  Phase kickback is implemented here with a ripple-carry adder~\cite{Cuccaro2004}.  PARs use six pre-computed ancillas.  Solovay-Kitaev sequences were calculated using code written by Dawson~\cite{Dawson05}; Fowler sequences were calculated using code written by Fowler.}
  \label{Single_qubit_phase_depth}
\end{figure}

\begin{table}
  \centering
  \begin{tabular}{|>{\raggedright\arraybackslash}m{2.3cm}|>{\raggedright\arraybackslash}m{3cm}|>{\raggedright\arraybackslash}m{3cm}|>{\raggedright\arraybackslash}m{3cm}|}
    \hline
    \textbf{Method} & \textbf{Description} & \textbf{Advantages} & \textbf{Disadvantages} \\ \hline
     Phase kickback & Approximates arbitrary phase rotation via controlled addition applied to $\ket{\gamma^{(k)}}$ ancilla register. & Trivial to compile.  Circuit depth is $O(\log \epsilon)$ or $O(\log \log \epsilon)$, depending on adder circuit. & Requires a logical ancilla register consisting of $O(\log \epsilon)$ qubits.  Resource costs are about 2--3$\times$ higher than Fowler sequences. \\ \hline
     Solovay-Kitaev sequence & Approximates arbitrary rotation with a sequence of fundamental gates.  Depth is $O(\log^c \epsilon)$, with $c \approx 4$.  & Polynomial-time compiling algorithm. No logical ancilla states. & Dramatically more expensive in quantum resources than alternatives.\\ \hline
     Fowler sequence & Approximates arbitrary rotation with a sequence of fundamental gates.  Depth is $O(\log \epsilon)$.  & Minimal-depth sequences.  No logical ancilla states. & Sequence-determination algorithm has exponential complexity and becomes infeasible for high-accuracy rotations. \\ \hline
     Programmable ancilla rotation (PAR) & Approximates arbitrary rotation with a probabilistic circuit using ancilla and measurement. & Constant average depth (4 gates) for any phase rotation. & Requires logical ancillas which must be pre-computed. \\ \hline
  \end{tabular}
  \caption{Summary of methods for producing fault-tolerant phase rotations.  The quantity $\epsilon$ is the accuracy of an approximate rotation, and it is defined by Eqns.~(\ref{distance_metric}) and~(\ref{define_epsilon}).}
  \label{phase_rotations_summary}
\end{table}

When benchmarking the performance of a phase rotation, the important figures are the quantum resources consumed to achieve a given accuracy of approximation.  Using the distance measure in Eqn.~(\ref{distance_metric}), the approximation error is quantified as
\begin{equation}
\epsilon = \mathrm{dist}_2\left(R_Z(\phi),U_{\mathrm{approx}}\right),
\label{define_epsilon}
\end{equation}
where $U_{\mathrm{approx}}$ is the circuit approximating $R_Z(\phi)$.  Figure~\ref{Single_qubit_phase_depth} reports two quantum resources for a single-qubit rotation: circuit depth, which is the minimum execution time in gates; and the total number of \texttt{T}~gates required (see Table~\ref{gate_set}).  \texttt{T}~gates are significantly more expensive to prepare fault-tolerantly than other fundamental gates in many prominent error-correcting codes~\cite{Nielsen00,Fowler09}, so they represent an important consideration for large-scale quantum computing~\cite{Isailovic08,Clark09,Jones2011}.  It is apparent from figure~\ref{Single_qubit_phase_depth} that Solovay-Kitaev sequences are substantially more expensive than their counterparts in both circuit depth and \texttt{T}~gates.  Fowler sequences are very compact and, in fact, optimal for an approximation sequence, but the classical algorithm to calculate them requires a calculation time that appears to grow exponentially faster than the other methods: $\epsilon \le 10^{-2}$ requires minutes, $\epsilon \le 10^{-3}$ requires about an hour, and $\epsilon \le 10^{-4}$ requires about a day, for each rotation, on a modern workstation.  For these reasons, phase kickback may be the method of choice when high-precision ($\epsilon \le 10^{-6}$) rotations are required.  Phase kickback requires resources comparable to Fowler sequences, but the quantum circuit depends on adders, which are trivial to compile.  The methods we analyze for producing fault-tolerant phase rotations are summarized in Table~\ref{phase_rotations_summary}.

\section{Simulating chemistry in second-quantized representation}
\label{second_quantized}
Simulation in the second-quantized form expresses the electronic Hamiltonian $\mathcal{H}$ in terms of the creation operators ${a_p}^ \dag$ and the wavefunction in terms of fermionic (or bosonic) modes $\left| p \right\rangle  \equiv {a_p}^\dag \left| 0 \right\rangle$ (\emph{i.e.}, occupation number representation).  In chemistry, the single-electron molecular orbital picture has provided a practical method for approximating an $N$-electron wavefunction.  Using second-quantized algorithms, basis sets in computational chemistry can be imported directly into quantum computational algorithms.  For this reason, both theoretical~\cite{Aspuru05,wang2008quantum,Whitfield2011} and experimental~\cite{Lanyon2010,Du2010} investigations in second-quantization have been performed.

\begin{figure}
  \centering
  \includegraphics[width=\textwidth]{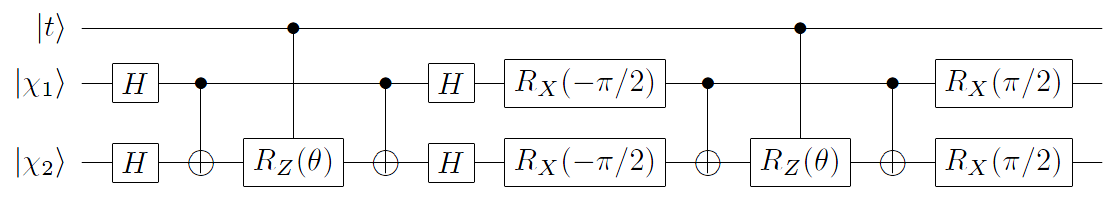}\\
  \caption{Excitation operator $e^{-ih_{12}({a_1}^{\dag} a_{2}  + {a_2}^{\dag} a_{1})\delta t}$ encoded into a quantum circuit~\cite{Whitfield2011}.  Above, $\theta = h_{12} \delta t$.  The gate $R_X(-\pi/2) = \texttt{H} \cdot \texttt{S}^{\dag} \cdot \texttt{H}$ is available from the set in Table~\ref{gate_set}.  In this example, the control qubit $\ket{t}$ is used for phase estimation, and the qubits $\ket{\chi_1}$ and $\ket{\chi_2}$ are basis functions (\emph{e.g.} molecular orbitals).  The controlled phase rotations $CR_Z(\theta)$ must be approximated using circuits of available fault-tolerant gates.}
  \protect\label{excitation_operator}
\end{figure}

Following the standard construction (see \emph{e.g.} Ref.~\cite{Kassal2011}), an arbitrary molecular Hamiltonian in second-quantized form can be expressed as
\begin{equation}
\label{second_hamiltonian}
\mathcal{H} = \sum_{p,q}h_{pq}{a_p}^{\dag}a_q + \frac{1}{2}\sum_{p,q,r,s}h_{pqrs}{a_p}^{\dag}{a_q}^{\dag}a_r a_s \; ,
\end{equation}
where $h_{pq} =\langle p | (\hat{T} + \hat{V}_N) | q \rangle$ are one-electron integrals ($\hat{T}$ is the kinetic energy operator, and $\hat{V}_N$ is the nuclear potential) and $h_{pqrs} = \langle pq | \hat{V}_e | rs \rangle$ represent the Coulomb potential interactions between electrons.  All of the terms $h_{pq}$'s and $h_{pqrs}$'s are pre-computed numerically with classical computers, and the values are then used in the quantum computer to simulate the Hamiltonian evolution through the operators
\begin{equation}
U_{pq} = e^{-ih_{pq}({a_p}^{\dag}a_q+{a_q}^{\dag}a_p) \delta t}
\end{equation}
and
\begin{equation}
\label{two_body_propagator}
U_{pqrs} = e^{-ih_{pqrs}({a_p}^{\dag}{a_q}^{\dag}a_r a_s + {a_s}^{\dag}{a_r}^{\dag}a_q a_p)\delta t}.
\end{equation}
These operators are constructed with a Jordan-Wigner transform and an arbitrary controlled phase gate $CR_Z(\phi)$~\cite{Whitfield2011}, as shown in figure~\ref{excitation_operator}.  The Jordan-Wigner transform requires \texttt{H}, \texttt{S}, and \texttt{CNOT} gates, which are often readily available in fault-tolerant settings, so we focus first on the considerably more resource-intensive controlled phase rotations.  We later show how to implement the Jordan-Wigner transform efficiently.

\subsection{Controlled phase rotations}
\label{controlled_rotations}
\begin{figure}
  \centering
  %
  %
  %
  %
  %
  \includegraphics[width=\textwidth]{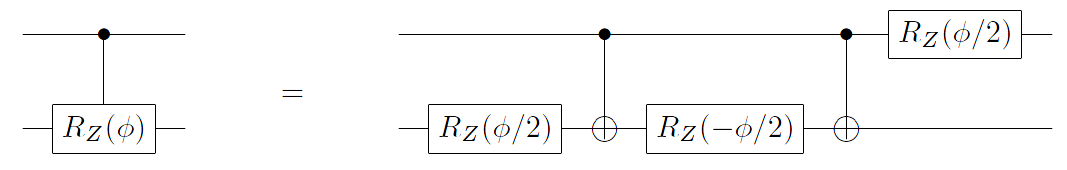}\\
  \caption{Decomposition of a controlled phase rotation into \texttt{CNOTs} and fault-tolerant single-qubit rotations.  If the control qubit \emph{only controls other circuits}, as in phase estimation algorithms, the third phase rotation commutes with the \texttt{CNOTs}.  In such an event, the third single-qubit rotations from all decompositions of controlled rotations commute, and they can be combined into just one rotation prior to a non-commuting operation on this qubit (such as the quantum Fourier transform and measurement readout in figure~\ref{sim_algorithm}).  As a result, controlled rotations in phase estimation algorithms are effectively decomposed into two \texttt{CNOTs} and two single-qubit rotations with this circuit.}
  \protect\label{controlled_rotation_figure}
\end{figure}

\begin{figure}
  \centering
  %
  %
  %
  %
  %
  \includegraphics[width=9cm]{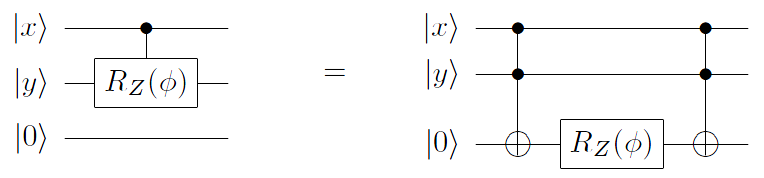}\\
  \caption{Controlled rotation $CR_Z(\phi)$ (see Eqn.~(\ref{controlled_rotation_eqn})) between qubits $\ket{x}$ and $\ket{y}$ using two Toffoli gates, just one single-qubit rotation gate, and an ancilla $\ket{0}$.  The ancilla qubit is conditionally set to $\ket{1}$ using a Toffoli gate, and a phase is imparted to this state with the rotation $R_Z(\phi)$.  A final Toffoli gate returns the ancilla qubit to state $\ket{0}$.}
  \protect\label{fast_controlled_rotation}
\end{figure}

As can be seen in figure~\ref{excitation_operator}, when $U_{pq}$ or $U_{pqrs}$ is implemented in a controlled operation (such as in energy eigenvalue estimation, see also figure~\ref{sim_algorithm}), the core component of the circuit is a controlled phase rotation,
\begin{equation}
\label{controlled_rotation_eqn}
CR_Z(\phi) = \left[\begin{array}{cccc} 1 & 0 & 0 & 0 \\ 0 & 1 & 0 & 0 \\ 0 & 0 & 1 & 0 \\ 0 & 0 & 0 & e^{i\phi} \end{array} \right].
\end{equation}
One way to implement the controlled rotation in Eqn.~(\ref{controlled_rotation_eqn}) is to deconstruct the operation into \texttt{CNOTs} and single-qubit rotations~\cite{Fowler2007}, as shown in figure~\ref{controlled_rotation_figure}.  Another method requires just one single-qubit rotation, as well as an ancilla $\ket{0}$, as shown in figure~\ref{fast_controlled_rotation}.  Ref.~\cite{Nielsen00} provides a circuit decomposition for the Toffoli gate into gates in Table~\ref{gate_set}.  We use the circuit in figure~\ref{fast_controlled_rotation} (requiring just one phase rotation) for the remainder of this paper, because the cost of one ancilla qubit is typically modest compared to a phase rotation.  One can implement phase kickback, gate approximation sequences, or PARs to produce the single-qubit rotations, as in Section~\ref{single_qubit_rotations}.  Additionally, the PAR construction can be modified to produce controlled rotations more directly.  If the control qubit \emph{only controls other circuits} between ancilla production and the time a controlled-PAR is needed, as is the case for phase estimation algorithms, one can create the ancillas (see figure~\ref{PAR_circuit}) using controlled rotations with one of the above methods and produce a controlled-PAR with the same cascading circuit.

\begin{figure}
  \centering
  \includegraphics[width=14cm]{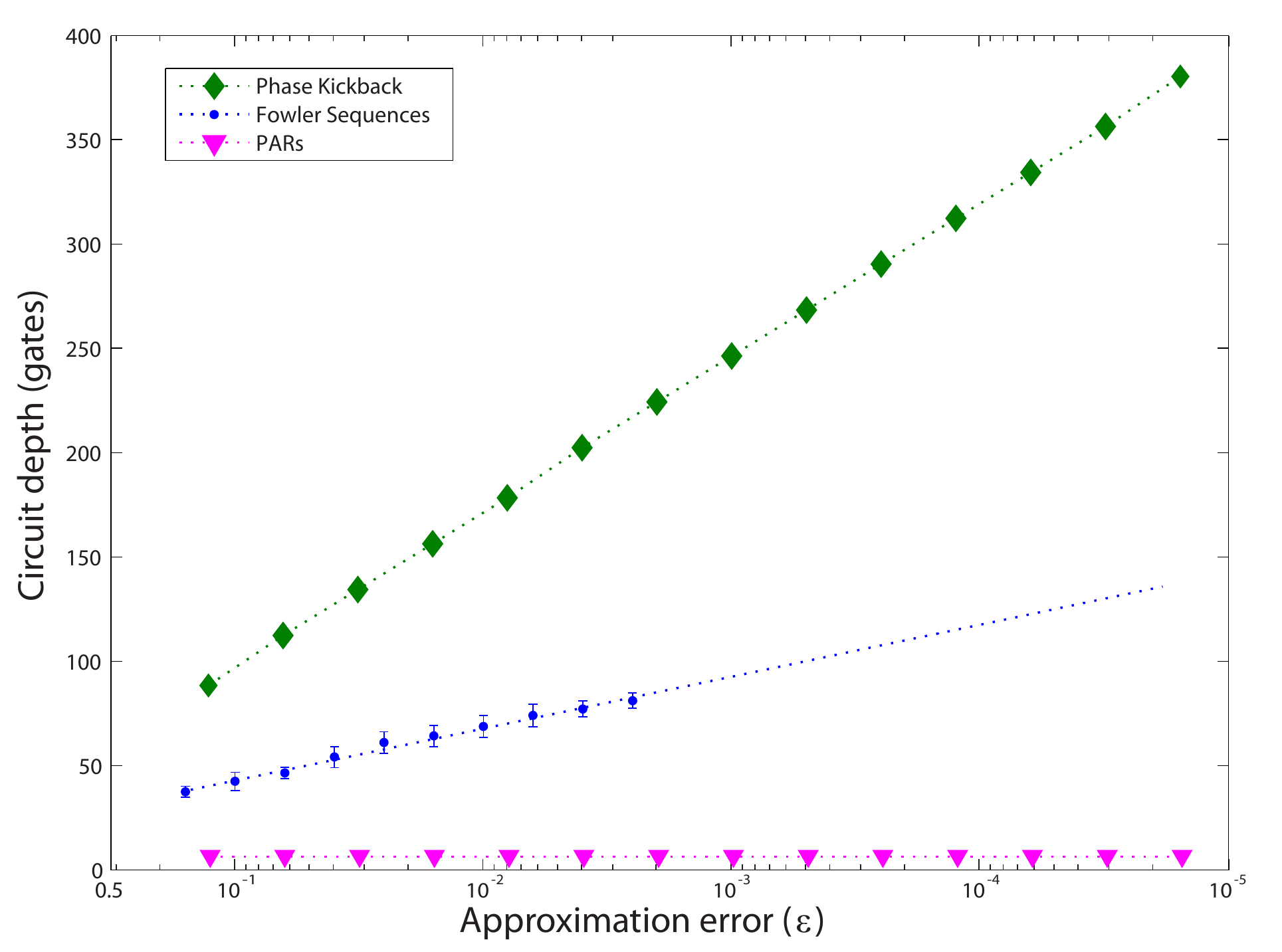}\\
  \caption{Color.  Circuit depth for controlled phase rotations using various methods.  A desired controlled rotation $CR_Z(\phi)$ is approximated with a fault-tolerant circuit $U_{\mathrm{approx}}$ with accuracy $\epsilon = \mathrm{dist}_4\left(U_{\mathrm{approx}},CR_Z(\phi)\right)$ using the method in figure~\ref{fast_controlled_rotation}.  Solovay-Kitaev sequences are omitted here to permit comparison of the more efficient schemes on a linear scale.  The bars on Fowler sequence data indicate the standard deviation taken over 98 random-angle rotations.  The controlled-PARs have depth of 4 gates, on average, regardless of rotation accuracy.  Phase kickback uses a ripple-carry adder since the addends have less than 16 bits~\cite{Cuccaro2004}.  If very high precision were desired, a carry-lookahead adder can achieve depth $O(\log \log \epsilon)$ at the expense of additional qubits and parallel circuits (more \texttt{T} gates)~\cite{Draper2006}.}
  \label{CRZ_depth}
\end{figure}

The different methods of producing a controlled phase rotation are analyzed in figure~\ref{CRZ_depth}.  We have excluded Solovay-Kitaev sequences, which permits a linearly-scaled vertical axis, showing that each of these methods has execution time linear in $\log \epsilon$ or constant.  As before, the values for Fowler sequences are extrapolated.  We can see that Fowler sequences and phase kickback are separated by approximately a factor of 3 in execution time, and the choice between the two would be motivated by whether compiling the Fowler sequence is feasible or not.  The PAR circuit requires one of the above methods to pre-compute ancillas.

\subsection{Finite precision in pre-calculated integrals}
\label{finite_integrals}
The execution time of a second-quantized simulation algorithm is proportional to the number of integral terms $h_{pq}$ and $h_{pqrs}$, as indicated by Eqns.~(\ref{second_hamiltonian}--\ref{two_body_propagator}).  We now consider how to speed up the algorithm by omitting the integral terms that are negligibly small in magnitude.  For a basis set consisting of $M$ single-particle orbitals, the maximum number of integral terms is $O(M^4)$.  In practice, however, the effort for evaluating these integrals often scales somewhere between $O(M^2)$ and $O(M^3)$ with modern implementations~\cite{Helgaker2000}, because typically many integral terms may be neglected for being smaller in magnitude than a cutoff threshold.  Consequently, the execution time of second-quantized simulation is determined by the number of pre-computed integrals of the form $h_{pq}$ and $h_{pqrs}$ of sufficiently large magnitude, as well as the efficiency of producing the corresponding arbitrary phase rotations in the quantum computer, such as $CR_Z(h_{pq} \delta t)$ in the gate sequence for $e^{-ih_{pq}(a_p^{\dag}a_q+a_q^{\dag}a_p) \delta t}$~\cite{Whitfield2011}.

\begin{figure}
\centering
\includegraphics[width=\textwidth]{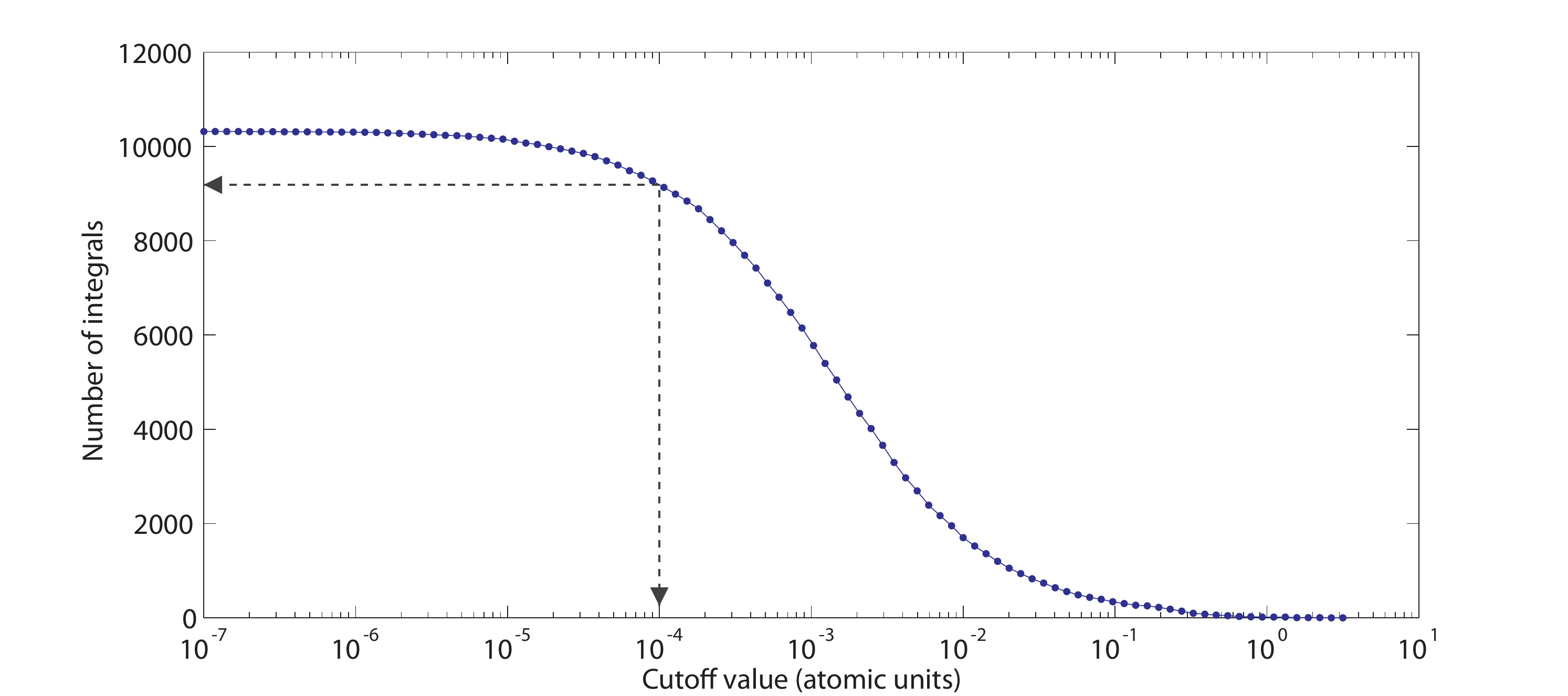}
\caption{Color.  The number of integral terms implemented in a second-quantized simulation of LiH using a TZVP basis, as a function of cutoff threshold.  Only integral terms with absolute value above the threshold are implemented in circuits, and the rest are neglected.  As shown in the figure, a cutoff of $10^{-4}$ would require the algorithm to implement just over 9000 integral terms.}
\label{fig:ints}
\end{figure}

To illustrate how many integral terms are present in a typical chemical problem, we have calculated the integrals for a second-quantized simulation of LiH.  We performed calculations in the minimal basis and in a triple-zeta basis, using the GAMESS quantum chemistry package~\cite{GAMESS1,GAMESS2}, at a bond distance of 1.63 \AA, with an integral term cutoff of $10^{-10}$ in atomic units.  We computed the number of integrals above cutoff using the STO-3G basis~\cite{Hehre69} containing 12 spin-orbitals (6 spatial orbitals) and the TZVP basis~\cite{Dunning71} containing 40 spin orbitals (20 spatial orbitals).  The cumulative number of integral terms as a function of cutoff in TZVP basis is plotted in figure~\ref{fig:ints}.  With the STO-3G basis, there were 231 non-zero molecular integrals, but only 99 of them were greater than 10$^{-10}$ atomic units in magnitude.   This is an order of magnitude below what is expected from $O(M^4)$ scaling.  Considering the larger, more accurate basis set (TZVP), there were 22155 non-zero integrals, but only 10315 were greater than the cutoff.  Figure~\ref{fig:ints} shows that a higher cutoff, such as $10^{-4}$, can further reduce the number of integrals in TZVP basis implemented in the simulation.  As a result, the effective number of integral terms the quantum computer must implement as phase rotations is nearly two orders of magnitude less than the asymptotic analysis would suggest, an example of the over-estimation of the resource costs that can occur when using asymptotic estimates.  This technique becomes particularly relevant in large molecules since distant particles interact weakly, and in such an event, many of the associated integral terms may be negligibly small.  Raising the cutoff threshold impacts the accuracy of the simulation, so one must attempt to balance the resource costs of simulation with the usefulness of the result.

\subsection{Jordan-Wigner transform using teleportation}
\label{teleport_JW}
\begin{figure}
\centering
\includegraphics[width=\textwidth]{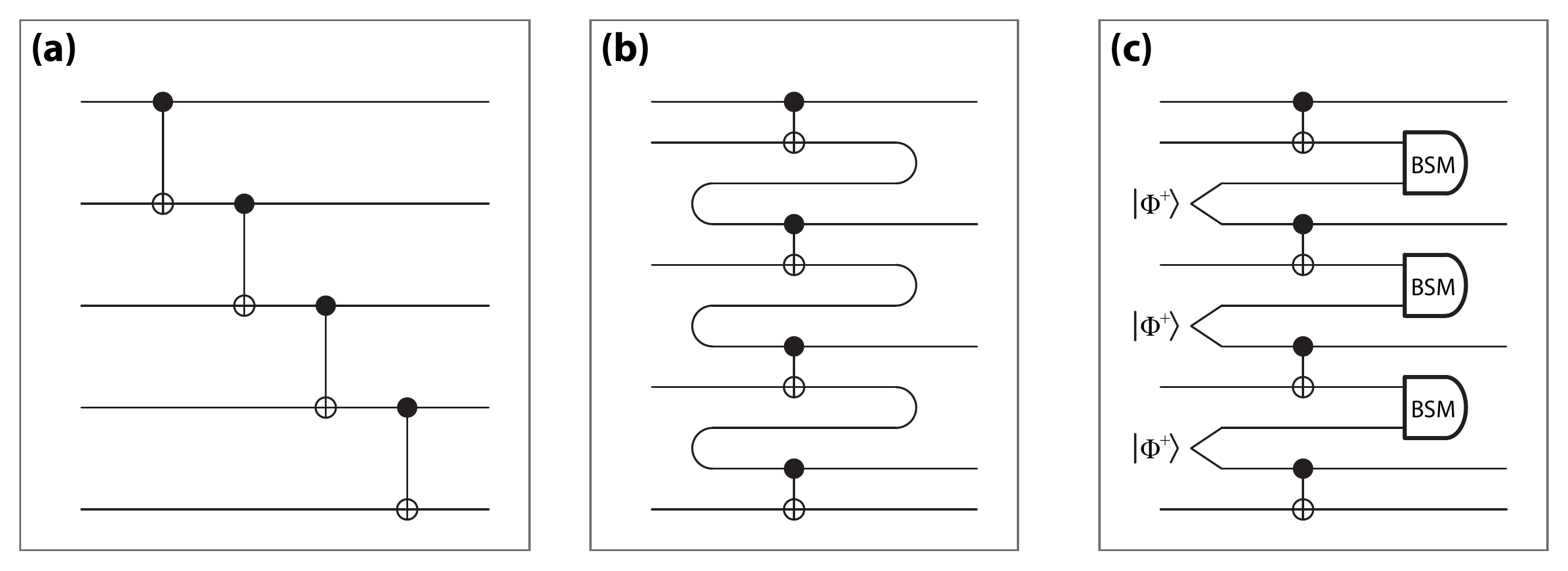}
\caption{Rearrangement of the \texttt{CNOT} ladder common in Jordan-Wigner transforms using teleportation.  \textbf{(a)} The original \texttt{CNOT} ladder requires an execution time that grows with the extent of the simulation in qubits.  \textbf{(b)} A conceptual diagram of what teleportation accomplishes.  The qubits ``move'' backwards in time.  \textbf{(c)} A valid quantum circuit that uses teleportation to move qubits in a manner which allows parallel computation of the \texttt{CNOTs}.  The \texttt{BSM} is the Bell state measurement which teleports the qubits; the result of this measurement indicates the Pauli errors which are tracked by the Pauli frame~\cite{Jones2011}.  The Bell state $|\Phi^{+}\rangle = \frac{1}{\sqrt{2}}\left(\ket{00} + \ket{11}\right)$ can be prepared from $\ket{0}$ ancillas using one \texttt{H} gate and one \texttt{CNOT} gate.  Similarly, the \texttt{BSM} can be implemented using one \texttt{H}, one \texttt{CNOT}, and measurement of the two qubits in the computational basis.}
\label{JWTeleport}
\end{figure}

The second-quantized algorithm uses Jordan-Wigner transforms to implement operators such as $e^{-ih_{pq}({a_p}^{\dag}a_q+{a_q}^{\dag}a_p) \delta t}$, and this section shows how to perform such transforms in constant time.  As elaborated in Ref.~\cite{Whitfield2011}, the circuits for Jordan-Wigner transforms often consist of ladders of \texttt{CNOT} gates, such as the one in figure~\ref{JWTeleport}a.  In a simulation with $M$ basis states, these ladders can extend across the entire register of qubits corresponding to these basis states, which leads to the $O(M^5)$ asymptotic runtime quoted in Ref.~\cite{Kassal2011} when there are at most $O(M^4)$ integral terms.

The \texttt{CNOT} ladder is a sparse network of Clifford gates, so we show how it may be implemented in constant time using teleportation~\cite{Gottesman1999,Zhou2000}.  Figure~\ref{JWTeleport}b gives an intuitive picture for what will be accomplished.  If the path of the qubits could be rearranged to somehow propagate backwards in time, the \texttt{CNOT} gates could be implemented simultaneously.  Qubits cannot move backwards in time \emph{per se}, but they can be moved arbitrarily using teleportation; notice how the conceptual (but unphysical) circuit in figure~\ref{JWTeleport}b is realized by a physical circuit in figure~\ref{JWTeleport}c.  Ancilla Bell states $|\Phi^{+}\rangle = \frac{1}{\sqrt{2}}\left(\ket{00} + \ket{11}\right)$ are used to teleport qubits in this rearranged \texttt{CNOT} ladder.  Teleportation introduces a random Pauli error on the teleported qubit, but it is possible to track these errors and their propagation through \texttt{CNOT} gates using Pauli frames~\cite{Knill05,DiVincenzo07,Jones2011}.  With this modification, it is possible to implement the Jordan-Wigner transform in constant time, which removes one of the bottlenecks to high-speed second-quantized simulation.  This method could be adapted to implement other Clifford-group circuits in constant time, at the expense of requiring enough ancilla Bell states.

\subsection{Resource analysis for ground-state energy simulation of LiH}
\begin{figure}
  \centering
  \includegraphics[width=\textwidth]{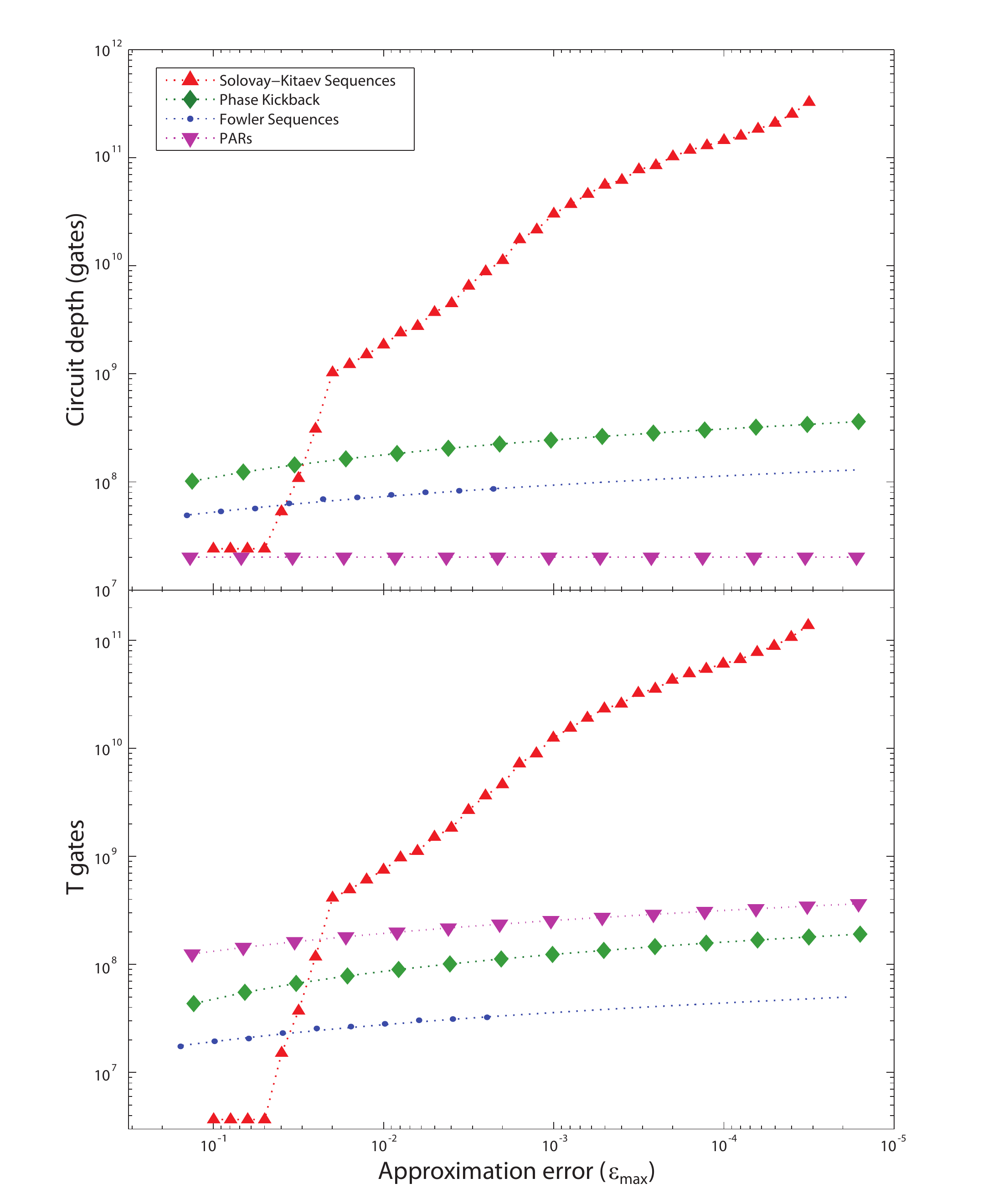}\\
  \caption{Color.  Total circuit depth and \texttt{T} gates for a second-quantized simulation of LiH using the STO-3G basis, calculated for different constructions of controlled rotations as a function of accuracy $\epsilon_{\mathrm{max}}$.  For a given $\epsilon_{\mathrm{max}}$, every controlled rotation $CR_Z(\phi)$ in the algorithm is approximated with a fault-tolerant circuit $U_{\mathrm{approx}}$ with accuracy distance $\epsilon = \mathrm{dist}_4\left(U_{\mathrm{approx}},CR_Z(\phi)\right)$ such that $\epsilon \le \epsilon_{\mathrm{max}}$.  An accuracy threshold $\epsilon_{\mathrm{max}} \le 10^{-4}$ is used in later analysis.  This simulation implements all integral terms in the Hamiltonian (see Eqn.~(\ref{second_hamiltonian})).  \textbf{(top)} Circuit depth using the gate set in Table~\ref{gate_set}.  In this plot, only the mean number of gates for PAR circuits is shown.  \textbf{(bottom)} \texttt{T} gates required for each method.  The controlled-PAR ancillas are produced using controlled rotations constructed using Fowler sequences; 6 controlled-PAR ancillas are pre-computed for each rotation, and only mean values are plotted.  The sudden jump in Solovay-Kitaev resource costs is because many controlled rotations in this algorithm have a small angle $\phi \approx 0$ that is approximated with identity gate at low precision, whereas the other methods are using a typical sequence length for arbitrary $\phi$.}
  \label{LiH_combined}
\end{figure}

Using the hypothetical quantum computer from Section~\ref{single_qubit_rotations}, we examine the resources required to perform simulation in second-quantized form.  Estimates of the number of qubits required for various instances of second-quantized chemical simulation have been reported previously~\cite{Aspuru05,Kassal2011}, so we focus instead on the execution time and effort to prepare fault-tolerant gates (here we consider number of \texttt{T}~gates).  Figure~\ref{LiH_combined} shows both the circuit depth and number of \texttt{T}~gates required to simulate LiH in the STO-3G basis as a function of rotation accuracy threshold $\epsilon_{\mathrm{max}}$, for 1023 simulated time steps.  The precision in the readout is proportional the number of time steps simulated.  The energy estimate in this simulation has 10 bits of precision, and in general, $2^n - 1$ steps are required for $n$ bits of precision.  If we assume that the duration of a single quantum gate is 1 ms (\emph{cf.} Ref.~\cite{Jones2011}), then the total execution time of the simulation ranges from $\sim 5.6$ hours using PARs to $\sim 3.8$ years using Solovay-Kitaev rotations.

\begin{figure}
  \centering
  \includegraphics[width=\textwidth]{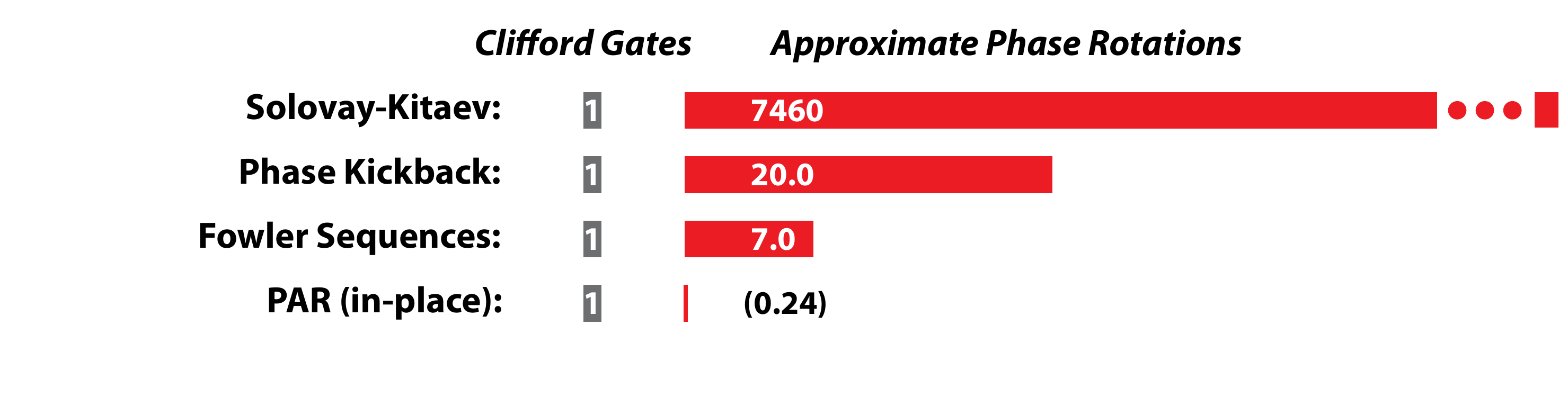}\\
  \caption{Color.  The relative amount of time (circuit depth) of a fault-tolerant, second-quantized simulation of LiH devoted to Clifford gates \{\texttt{X},\texttt{Y},\texttt{Z},\texttt{H},\texttt{S},\texttt{CNOT}\} versus phase rotations that must be approximated.  In this example, rotations are computed to an accuracy $\epsilon \le 10^{-4}$.  The relative circuit depth of rotations calculated by the Solovay-Kitaev algorithm is too large to be drawn to scale here.  In the case of PAR, the ancillas must be pre-computed with a method such as Fowler sequences, but this can be carried out in parallel with other algorithm operations.}
  \label{Relative_depth_rotations_LiH}
\end{figure}

The number of \texttt{T}~gates in figure~\ref{LiH_combined} serves as an indication of the complexity demanded of the quantum computer.  Although we do not delve into this matter, Refs.~\cite{Isailovic08,Jones2011} discuss the importance (and difficulty) of producing these gates.  What becomes apparent is that using PARs, while very fast, is also more expensive in the consumption of \texttt{T}~gates than directly implementing Fowler sequences or phase kickback.  Choosing between such approaches depends on the capabilities of the quantum computer, and we discuss this matter in more detail in Section~\ref{comparing_methods}.

\begin{table}
  \centering
  \begin{tabular}{|>{\raggedright\arraybackslash}m{2.3cm}|>{\raggedright\arraybackslash}m{3cm}|>{\raggedright\arraybackslash}m{3cm}|>{\raggedright\arraybackslash}m{3cm}|}
    \hline
    \textbf{Method} & \textbf{Description} & \textbf{Advantages} & \textbf{Disadvantages} \\ \hline
     Finite-precision cutoff in second-quantized integrals & Neglect to implement integral terms below a chosen cutoff in the algorithm execution. & Second-quantized circuit complexity is reduced in both depth and number of \texttt{T}~gates. & None if cutoff threshold is below gate approximation accuracy. \\ \hline
     Jordan-Wigner transform using teleportation & Use a teleportation circuit to implement Jordan-Wigner transform in constant time. & Second-quantized circuit depth reduces to at most $O(M^4)$ from $O(M^5)$. & Teleportation circuit requires at most $3M-4$ qubits instead of $M$ (only during Jordan-Wigner transform). \\ \hline
  \end{tabular}
  \caption{Summary of methods for efficient second-quantized chemical simulation.  The quantity $M$ is the number of basis functions used in the representation of the chemical problem; larger basis sets produce more accurate results at the expense of greater circuit complexity.}
  \label{second_quant_summary}
\end{table}

To provide an indication of how much execution time in second-quantized simulation is devoted to phase rotations, figure~\ref{Relative_depth_rotations_LiH} shows the relative ratio of circuit depth devoted to implementing rotations versus all other gates for each of the methods considered when simulating LiH with rotation accuracy $\epsilon \le 10^{-4}$.  It is clear here that Solovay-Kitaev has such high circuit depth that it cannot be drawn to scale.  We see also that Fowler and phase kickback sequences require execution times that are comparable, whereas PARs actually do not represent the majority of the circuit depth, unlike all of the prior methods.  This is an encouraging result, because it shows that previous examinations that depended on Solovay-Kitaev sequences can be improved by orders of magnitude with more efficient phase rotations~\cite{Clark09}.  We do not consider Solovay-Kitaev sequences further in this investigation.  The techniques for improving second-quantized simulation are summarized in Table~\ref{second_quant_summary}.

\section{Simulating chemical structure and dynamics in first-quantized representation}
\label{first_quantized}
The first-quantized simulation algorithm is in some ways more complex than the second-quantized algorithm, but for problems in chemistry larger than a handful of particles, it is computationally faster.  A first-quantized simulation is essentially a finite-difference method for solving the Schr\"{o}dinger equation.  Configuration space is discretized into a Cartesian grid, and each particle (\emph{e.g.} electron) has a wavefunction expressed in a quantum register that which encodes a probability amplitude at each coordinate on the grid.  For example, let us imagine that we form a position-basis representation for a single electron on a $2^{p} \times 2^{p} \times 2^{p}$ grid, which requires only $3p$ qubits.  Explicitly, the electronic wavefunction is represented as
\begin{equation}
\label{first_quant_wavefunction}
\left|\psi_e\right\rangle = \sum_{x,y,z = 0}^{2^p - 1} c(x,y,z)\left|x\right\rangle\left|y\right\rangle\left|z\right\rangle = \sum_{\bf r} c({\bf r})\ket{\bf r},
\end{equation}
where $c(x,y,z)$ is the complex probability amplitude for the electron to occupy the volume element centered at the position ${\bf r} \equiv (x,y,z)$.  The rightmost part of Eqn.~(\ref{first_quant_wavefunction}) is shorthand that will be used throughout this section.  The spin degree of freedom can easily be incorporated by including an extra qubit, and to describe a many-electron state, the wavefunction has to be properly anti-symmetrized~\cite{Abrams1997,Ward2009a}.

To simulate the evolution of a time-independent molecular Hamiltonian $\mathcal{H}$ for problems in quantum chemistry, we adopt the method given in Refs.~\cite{Zalka1998b, Kassal2008}.  The complete Hamiltonian in first-quantized form can be expressed as the sum of the kinetic ($\hat{T}$) and potential ($\hat{V}$) operators
\begin{equation}\label{Hamiltonian}
\mathcal{H} = \hat{T} + \hat{V} =  - \sum\limits_i {\frac{{\hbar ^2 \nabla _i^2 }}{{2m_i }}}  + \frac{1}{2}\sum\limits_{i \ne j} {\frac{{q_i q_j }}{{4\pi \epsilon _0 r_{ij} }}},
\end{equation}
where the indices $i$ and $j$ run over all particles (electrons and nuclei) of any given molecule. Here $r_{ij}  \equiv \left| {{\bf r}_i  - {\bf r}_j } \right|$ is the distance between particles $i$ and $j$, which carry charges $q_i$ and $q_j$ respectively.

Let us outline how first-quantized simulation works before delving into details.  The core of the algorithm is evolving the Hamiltonian in simulated time, achieved by applying the propagator $\mathcal{U}(t)=\exp(-i\mathcal{H}t)$ (setting $\hbar = 1$ and assuming $\mathcal{H}$ is time-independent), which solves the time-dependent Schr\"{o}dinger equation~\cite{Lloyd1996}. This process is readily achieved using the split operator approximation, a form of Trotter-Suzuki decomposition~\cite{Suzuki1992,Brown2010,Wiebe2010,Kassal2011}, where the kinetic and potential energy operators are simulated in alternating steps as
\begin{equation}
\mathcal{U}(t) = e^{-i\mathcal{H}t} \approx \left[e^{-i \hat{T} \delta t/2}e^{-i \hat{V} \delta t}e^{-i \hat{T} \delta t/2}\right]^\frac{t}{\delta t}.
\end{equation}
The operators $e^{-i \hat{V} \delta t}$ and $e^{-i \hat{T} \delta t}$ are diagonal in the position and momentum bases, respectively.  One can switch the encoded configuration space representation between these two bases by applying the quantum Fourier transform to each spatial dimension of the wavefunction (\emph{cf.} Eqn.~(\ref{first_quant_wavefunction})), which can be efficiently implemented in a quantum computer~\cite{Weinstein2001}.  Ref.~\cite{Kassal2008} shows how to construct quantum circuits for operators $e^{-i \hat{V} \delta t}$ and $e^{-i \hat{T} \delta t}$, and in this section, we complement that work with analysis of fault-tolerant versions of these operators.

To make an algorithm fault-tolerant, its constituent operations must be decomposed into circuits of fault-tolerant primitive gates such as those in Table~\ref{gate_set}.  Consider the potential energy propagator $e^{-i \hat{V} \delta t}$ as an example. Given a $b$-particle wavefunction in the position basis as
\begin{equation}
\ket{\psi_{1,2,...,b}} = \sum_{{\bf r}_1,{\bf r}_2,...,{\bf r}_b} c({\bf r}_1,{\bf r}_2,...,{\bf r}_b)\ket{{\bf r}_1{\bf r}_2...{\bf r}_b},
\end{equation}
where $c(\cdot)$ is the complex amplitude as a function of position in configuration space and subscripts correspond to particles in the system, one calculates the phase evolution of the potential operator $e^{-i\hat{V}\delta t}$ in three steps, as follows:
\begin{eqnarray}
\sum_{{\bf r}_1,...,{\bf r}_b} c({\bf r}_1,...,{\bf r}_b)\ket{{\bf r}_1...{\bf r}_b}\ket{000...} \nonumber \\
\longrightarrow \sum_{{\bf r}_1,...,{\bf r}_b} c({\bf r}_1,...,{\bf r}_b)\ket{{\bf r}_1...{\bf r}_b}\ket{V({\bf r}_1,...,{\bf r}_b)} \label{compute_V} \\
\longrightarrow \sum_{{\bf r}_1,...,{\bf r}_b} e^{-iV({\bf r}_1,...,{\bf r}_b)\delta t} c({\bf r}_1,...,{\bf r}_b)\ket{{\bf r}_1...{\bf r}_b}\ket{V({\bf r}_1,...,{\bf r}_b)} \label{first_quant_phase} \\
\longrightarrow \sum_{{\bf r}_1,...,{\bf r}_b} e^{-iV({\bf r}_1,...,{\bf r}_b)\delta t} c({\bf r}_1,...,{\bf r}_b)\ket{{\bf r}_1...{\bf r}_b}\ket{000...}. \label{uncompute_V}
\end{eqnarray}
First, Eqn.~(\ref{compute_V}) calculates the potential energy as a function of position coordinates~\cite{Kassal2008} (note that $\hat{V}$ is diagonal in this basis) and stores the result in a quantum register $\ket{V({\bf r}_1,{\bf r}_2,...,{\bf r}_b)}$ to some finite precision.  \ref{App_first_quant} describes how to implement this quantum circuit for molecular Hamiltonians.  Second, Eqn.~(\ref{first_quant_phase}) uses the $\ket{V({\bf r}_1,{\bf r}_2,...,{\bf r}_b)}$ register in a ``quantum variable'' phase rotation that imparts a phase to each grid point of the wavefunction in position basis proportional to the potential energy at those coordinates.  This section discusses how to implement the quantum variable rotation using fault-tolerant quantum circuits.  Finally, the quantum circuit from the first step is reversed in Eqn.~(\ref{uncompute_V}) to reset the $\ket{V({\bf r}_1,{\bf r}_2,...,{\bf r}_b)}$ register to $\ket{000...}$, also known as ``uncomputation''~\cite{Nielsen00}.  The sequence of these three steps is equivalent to the operation $e^{-i\hat{V}\delta t}\ket{\psi}$.

The kinetic energy propagator $e^{-i \hat{T} \delta t}$ is calculated similarly in three steps, with the second also being a quantum variable rotation.  This operator is diagonal in momentum basis, so we transform the representation of the system wavefunction from position basis $\{x,y,z\}$ to momentum basis $\{k_x,k_y,k_z\}$ by applying a QFT along each spatial dimension of the encoding in Eqn.~(\ref{first_quant_wavefunction}).  This form permits efficient calculation of the kinetic energy operator~\cite{Kassal2008}, which is described in \ref{App_first_quant}.

\subsection{Quantum variable rotation}
\label{QVR_section}
The phase rotation subroutine in the first-quantized simulation algorithm imparts a quantum phase to each binary-encoded phase state in a superposition $\ket{\theta} = \sum_j c_j\ket{\phi_j}$ stored in a quantum register ($c_j$'s are arbitary complex amplitudes).  Formally, it is the transformation
\begin{equation}
\label{quantum_variable_rotation}
\sum_j c_j\ket{\phi_j} \longrightarrow \sum_j e^{2\pi i \xi \phi_j} c_j\ket{\phi_j},
\end{equation}
which generalizes the operation in Eqn.~(\ref{first_quant_phase}) using $\xi$, which is a scaling factor that varies with implementation, as explained below and in \ref{App_first_quant}.  Each $0 \le \phi_j < 1$ is a finite binary representation of a rotation on the unit circle encoded in a quantum register.  Eqn.~(\ref{quantum_variable_rotation}) is the quantum variable rotation (QVR), which is essential to first-quantized simulation.  We show how to implement this phase rotation subroutine using phase rotations from previous sections, as well as a new construction based on phase kickback.  At the end of the section, we analyze the resource costs of these methods.

\begin{figure}
  \centering
  \includegraphics[width=5cm]{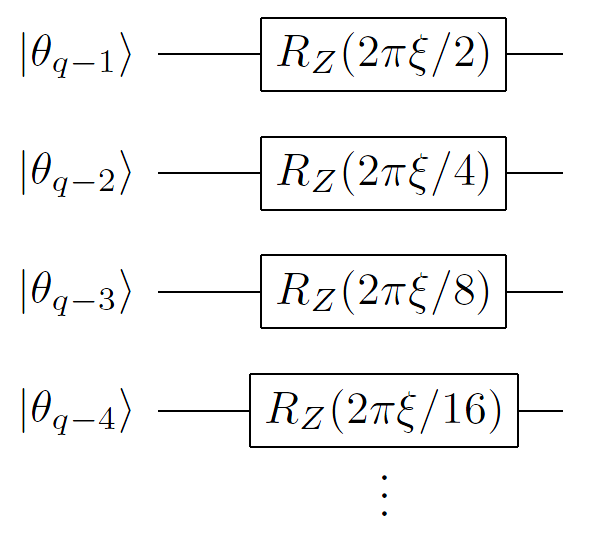}\\
  \caption{Quantum variable rotation decomposed into single-qubit rotations applied to each qubit in the $\ket{\theta}$ register consisting of $q$ qubits (see Eqn.~(\ref{quantum_variable_rotation})).  $\ket{\theta_{q-1}}$ refers to the most significant bit in the register $\ket{\theta}$, \emph{etc}.}
  \protect\label{QVR_bitwise}
\end{figure}

To produce a QVR, various circuit manipulations are possible.  The first is to simply apply a single-qubit rotation to each qubit in register $\ket{\theta}$, as shown in figure~\ref{QVR_bitwise}.  Each individual rotation could be created using the techniques in Section~\ref{phase_gates}.  Since a $t$-bit QVR requires $t$ separate bitwise rotations, we require that each rotation has accuracy $\epsilon/t$ to achieve accuracy $\epsilon$ in the QVR, where we have used the fact that the distance measure in Eqn.~(\ref{distance_metric}) obeys the triangle inequality~\cite{Fowler2011}.  If the QVR is controlled by another qubit (\emph{e.g.} if the propagator is controlled by a ``simulated time'' qubit as in figure~\ref{sim_algorithm}), then the gates in figure~\ref{QVR_bitwise} are replaced with controlled rotations from Section~\ref{controlled_rotations}.  In either case, one must know the quantity $\xi$ in advance to compile these gates; typically, $\xi$ is a product of physical constants and simulation parameters, as explained in \ref{App_first_quant}.

\begin{figure}
  \centering
  \includegraphics[width=10cm]{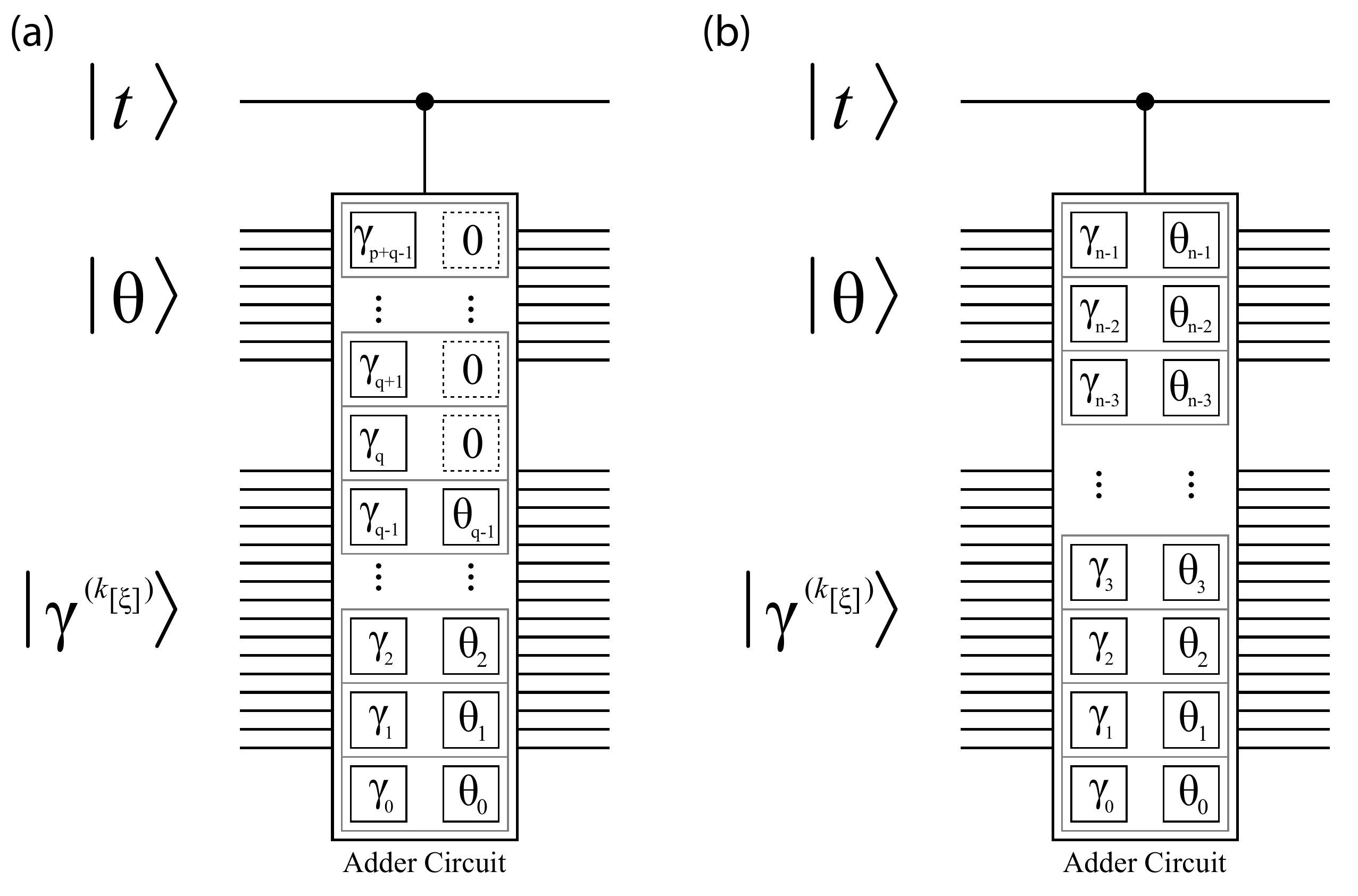}\\
  \caption{Quantum variable rotation using phase kickback.  This circuit implements the operation in Eqn.~(\ref{quantum_variable_rotation}) with scaling factor $[\xi]$, which has been ``programmed'' into the phase kickback register $\ket{\gamma^{(k_{[\xi]})}}$ (see \ref{transform_gamma}).  A control qubit $\ket{t}$ is included for illustration.  This figure shows how the bits in the adder are aligned for different cases. \textbf{(a)} The register $\ket{\theta}$ is shifted down $p$ bits since $p \ge 0$. $\theta_0$ is the least-significant bit in the $\ket{\theta}$ register, \emph{etc.}  The input qubits above $\ket{\theta}$ are logical zeros. \textbf{(b)} The register $\ket{\theta}$ is shifted up $|p|$ bits since $p < 0$.  In this case, the $|p|$ most-significant bits of $\ket{\theta}$ are not used in the adder.}
  \label{PK_QVR}
\end{figure}

The QVR can also be produced in a more elegant manner using phase kickback.  Rather than apply bitwise gates to the $\ket{\theta}$ register, we instead use the entire register in a modified version of the phase kickback procedure.  First, we require a binary approximation to $\xi$, denoted $[\xi]$.  Second, we define some quantities that describe this quantum circuit.  Let $m$ denote the number of significant bits in $[\xi]$, minus the number of trailing zeros.  Define $w = \lfloor \log_2 [\xi] \rfloor$, or in other words, $w$ is the largest integer such that $2^w \le [\xi]$.  Denote $p = (m-1) - w$, which is how many bits we must shift $[\xi]$ up to produce an odd integer (if $p < 0$, we shift down).  Following Eqn.~(\ref{quantum_variable_rotation}), let $q$ be the number of qubits in $\ket{\theta}$.  Define integers $k_{[\xi]} = (2^p) [\xi]$ and $u_{\phi} = (2^q) \phi$ for some arbitrary $\phi \in [0,1)$ represented using $q$ bits.  Third, we construct a phase kickback ancilla register $\ket{\gamma^{(k_{[\xi]})}}$ of size $n = p + q$ qubits, using techniques in \mbox{\ref{transform_gamma}}. Finally, we perform phase kickback with an addition circuit between registers $\ket{\theta}$ and $\ket{\gamma^{(k_{[\xi]})}}$ (in-place addition applied to $\ket{\gamma^{(k_{[\xi]})}}$), except this time the $\ket{\theta}$ register is shifted in one of two ways, as shown in figure~\ref{PK_QVR}.  If $p \ge 0$, then the $\ket{\theta}$ register is shifted down by $p$ qubits, and the $\ket{\theta}$ register is padded with $p$ logical zeros at the most-significant side of the adder input (figure~\ref{PK_QVR}a).  If $p < 0$, then $\ket{\theta}$ is shifted up by $|p|$ qubits, so that the $|p|$ most-significant bits of $\ket{\theta}$ are not used in the adder (figure~\ref{PK_QVR}b).  If $n \le 0$, then all rotations are identity and no QVR circuit is constructed.

\begin{figure}
  \centering
  \includegraphics[width=\textwidth]{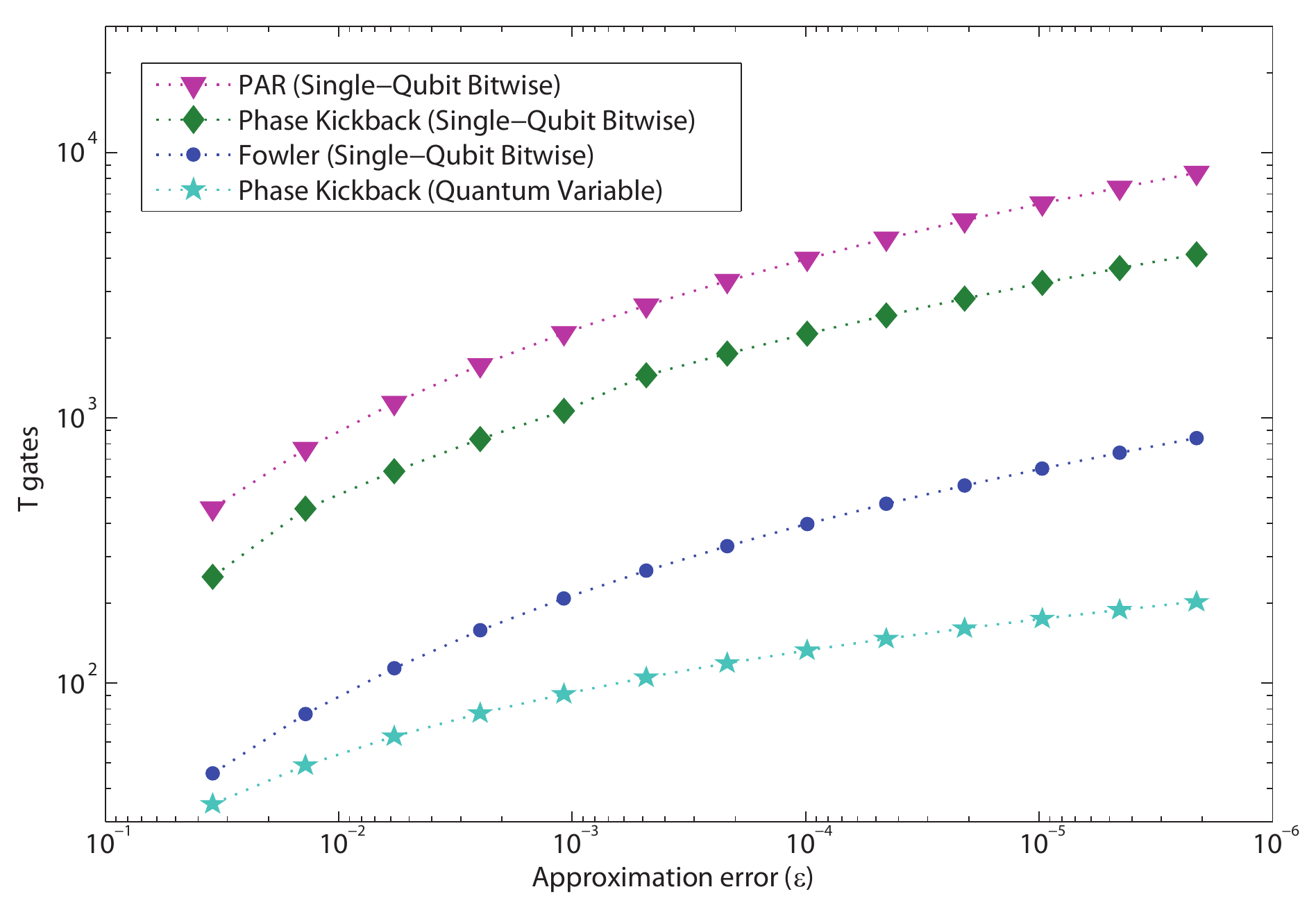}\\
  \caption{Color.  Number of \texttt{T} gates required to produce a QVR with various methods, assuming $\xi = 1$ and number of significant figures is chosen to satisfy the approximation error $\epsilon$.  The special-purpose ``quantum variable'' phase kickback clearly requires the least circuit effort, and the asymptotic scaling of \texttt{T} gates is linear in $\log \epsilon$ for this approach and super-quadratic for the others.  The circuit depth for Fowler or phase kickback approaches is equivalent to the comparable single-qubit rotation; however, the PAR must succeed across all individual rotations for this circuit to succeed, so the mean circuit depth increases slightly.  In the above, 10 rounds of PAR ancilla are pre-computed for each single-qubit rotation in the QVR.}
  \label{QVR_Tgates}
\end{figure}

We now confirm that this procedure produces the intended quantum variable rotation.  Using Eqn.~(\ref{PK_definition}), we see that the above procedure will implement a phase rotation of 
\begin{equation}
\sum_j c_j\ket{\phi_j} \longrightarrow \sum_j e^{2\pi i k_{[\xi]} u_{\phi_j}/2^{p+q}} c_j\ket{\phi_j}.
\end{equation}
Since $k_{[\xi]} = (2^p) [\xi]$ and $u_{\phi} = (2^q) \phi$, this is the same as
\begin{equation}
\sum_j c_j\ket{\phi_j} \longrightarrow \sum_j e^{2\pi i [\xi] \phi_j} c_j\ket{\phi_j},
\end{equation}
which is equivalent to Eqn.~(\ref{quantum_variable_rotation}) using our finite representation for $\xi$.  As before, if we require a controlled-QVR, then the adder can be controlled by an external qubit, which is the configuration shown in figure~\ref{PK_QVR}.  This ``quantum variable'' phase kickback uses substantially fewer \texttt{T} gates than the bitwise approach, as shown in figure~\ref{QVR_Tgates}, while having comparable circuit depth.  Moreover, since there is only one phase rotation instead of many, it does not have to be as accurate as the individual rotations in figure~\ref{QVR_bitwise} must be to achieve the same total accuracy in the QVR.

It may seem inefficient to produce a different phase kickback register for each QVR operation, but three properties of the first-quantized simulation algorithm make this approach efficient.  First, there are only a polynomial number of such operations: for $b$ particles, there are $b$ QVRs in the kinetic energy operator and $\frac{1}{2}b(b-1)$ QVRs in the potential operator.  Second, many of these QVRs have the same scaling factor $\xi$, so a phase kickback register can be reused many times without modification.  For example, the scaling factor in the kinetic energy operator is the same for all electrons (which have the same mass).  Third, the $\ket{\gamma^{(k_{[\xi]})}}$ registers can be calculated independently of other operations in the algorithm, so the impact of this process on circuit depth is minimal.

This phase kickback QVR has interesting applications to other useful quantum circuits.  It can be used to make a fault-tolerant quantum Fourier transform (QFT); one replaces each block of controlled rotations with a controlled-QVR.  As before, this approach uses substantially fewer \texttt{T} gates than an equivalent circuit where each controlled rotation in the QFT is implemented individually with techniques in Section~\ref{controlled_rotations}, and the same methods can be applied to an approximate QFT~\cite{Barenco1996} by simply truncating the size of the $\ket{\gamma^{(1)}}$ register.  The phase kickback QVR can also be used to efficiently produce ancillas for PAR if the particular rotation $R_Z(\phi)$ is required frequently, which can have applications to second-quantized simulation.  If we denote the state $\ket{+} = \frac{1}{\sqrt{2}}\left(\ket{0} + \ket{1}\right)$, then an input state of $\ket{+}\ket{+}\ket{+}...$ will be transformed using QVR (with appropriate $\xi$) into the set of ancillas for PAR, but requiring only one addition circuit for the entire set instead of a phase kickback addition or Fowler sequence for each ancilla qubit, which can be seen by comparing figure~\ref{QVR_bitwise} with the ancilla preparation in figure~\ref{PAR_circuit}.  Creating the necessary $\ket{\gamma^{(k_{[\xi]})}}$ for this process is costly, so there is a net gain only if a certain rotation angle $\phi$ is required often.

\subsection{Improved parallelism in potential energy operator}
The majority of the circuit effort in first-quantized simulation is devoted to calculating the potential energy~\cite{Kassal2008}.  We introduce here a technique to substantially speed up the calculation of the potential energy operator $\hat{V}$, which is simply the sum of the Coulomb interactions $\hat{V}_{ij} = \frac{q_i q_j}{4\pi \epsilon_0 r_{ij}}$ between all pairwise combinations of the electrons and nuclei.  Note that this operator is a function of the positions ${\bf r}_i$ of the system particles only, so it is diagonal in the position basis $\left| {{\bf r}_1 {\bf r}_2 ...{\bf r}_b } \right\rangle$.  This fact means that all terms $\hat{V}_{ij}$ commute with each other, so they may be calculated in any order.  Moreover, there are many sets of the $\hat{V}_{ij}$ operators that are disjoint, which means that each particle in the system is acted on by just one operator in the set.  Using this observation, for example, we may calculate the Coulomb interaction $\hat{V}_{12}$ between particles 1 and 2 at the same time as $\hat{V}_{34}$ between particles 3 and 4, and so on.  In general, for a system of $b$ particles, there are $\frac{1}{2}b(b-1)$ pairwise interactions, and we can perform $\lfloor \frac{b}{2} \rfloor$ pairs in parallel, which means that a potential energy operator with $O(b^2)$ terms can be calculated in $O(b)$ time.  This parallelism can increase the speed of simulation significantly since evaluation of the potential energy dominates resource costs~\cite{Jones2011}.

The potential operator calculation can be further parallelized to achieve $O(\log b)$ or $O(1)$ (constant) circuit depth.  Exploiting the fact that all $\hat{V}_{ij}$ are diagonal in position basis (and hence commute), we use transversal \texttt{CNOT} gates to copy the data in position-basis particle wavefunction onto multiple empty quantum registers.  For a single particle, this process is
\begin{eqnarray}
\left(\sum_{x,y,z = 0}^{2^p - 1} c(x,y,z)\left|x\right\rangle\left|y\right\rangle\left|z\right\rangle\right)\ket{000...}\ket{000...} ... \nonumber \\
\rightarrow \sum_{x,y,z = 0}^{2^p - 1} c(x,y,z)\left(\left|x\right\rangle\left|y\right\rangle\left|z\right\rangle\right) \left(\left|x\right\rangle\left|y\right\rangle\left|z\right\rangle\right) \left(\left|x\right\rangle\left|y\right\rangle\left|z\right\rangle\right) ...
\end{eqnarray}
For $b$ particles, the copy operation is performed $b-2$ times (for $b-1$ total copies), which can be fanned out using a binary tree with depth $\lceil\log_2(b-1)\rceil$; constant depth can be achieved in some quantum computer architectures which support one-control/many-target \texttt{CNOTs}~\cite{Fowler09,Jones2011} or in general architectures using a teleportation circuit similar to those described in Section~\ref{teleport_JW}.  This approach is similar to that employed in Ref.~\cite{Cleve2000} to produce a parallel circuit for the QFT.  The system wavefunction is now expanded to the state
\begin{equation}
\ket{\psi_{\mathrm{expand}}} = \sum_{{\bf r}_1,...,{\bf r}_b} c({\bf r}_1,...,{\bf r}_b)\left(\ket{{\bf r}_1}\right)^{\otimes (b-1)}...\left(\ket{{\bf r}_b}\right)^{\otimes (b-1)},
\end{equation}
which requires $O(b^2)$ memory space.  Note that this process is not cloning---the position-basis particle registers are still entangled to one another.  With multiple accessible copies of each particle's position-basis information, the particles are matched in all $b(b-1)$ possible pairings, and the potential energy operator applied to each pairing in parallel, which can be accomplished in constant time, but still requires $O(b^2)$ circuit effort.  After each of the potential energy operators $\hat{V}_{ij}$ kicks back a phase, the excess copies of each particle wavefunction are uncomputed by reversing the tree of \texttt{CNOTs} above.  The preceding example demonstrates that it is possible to calculate $\hat{V}$ in time which is sub-linear in the number of particles, even if each $\hat{V}_{ij}$ is treated as a black box operator.  In practice, more efficient circuits can be produced by generating the internal ``workspace'' registers of $\hat{V}$ in parallel, rather than making copies of the input registers $\sum_{{\bf r}_1,...,{\bf r}_b} c({\bf r}_1,...,{\bf r}_b)\ket{{\bf r}_1...{\bf r}_b}$ (see~\ref{App_first_quant}).

\begin{figure}
  \centering
  \includegraphics[width=\textwidth]{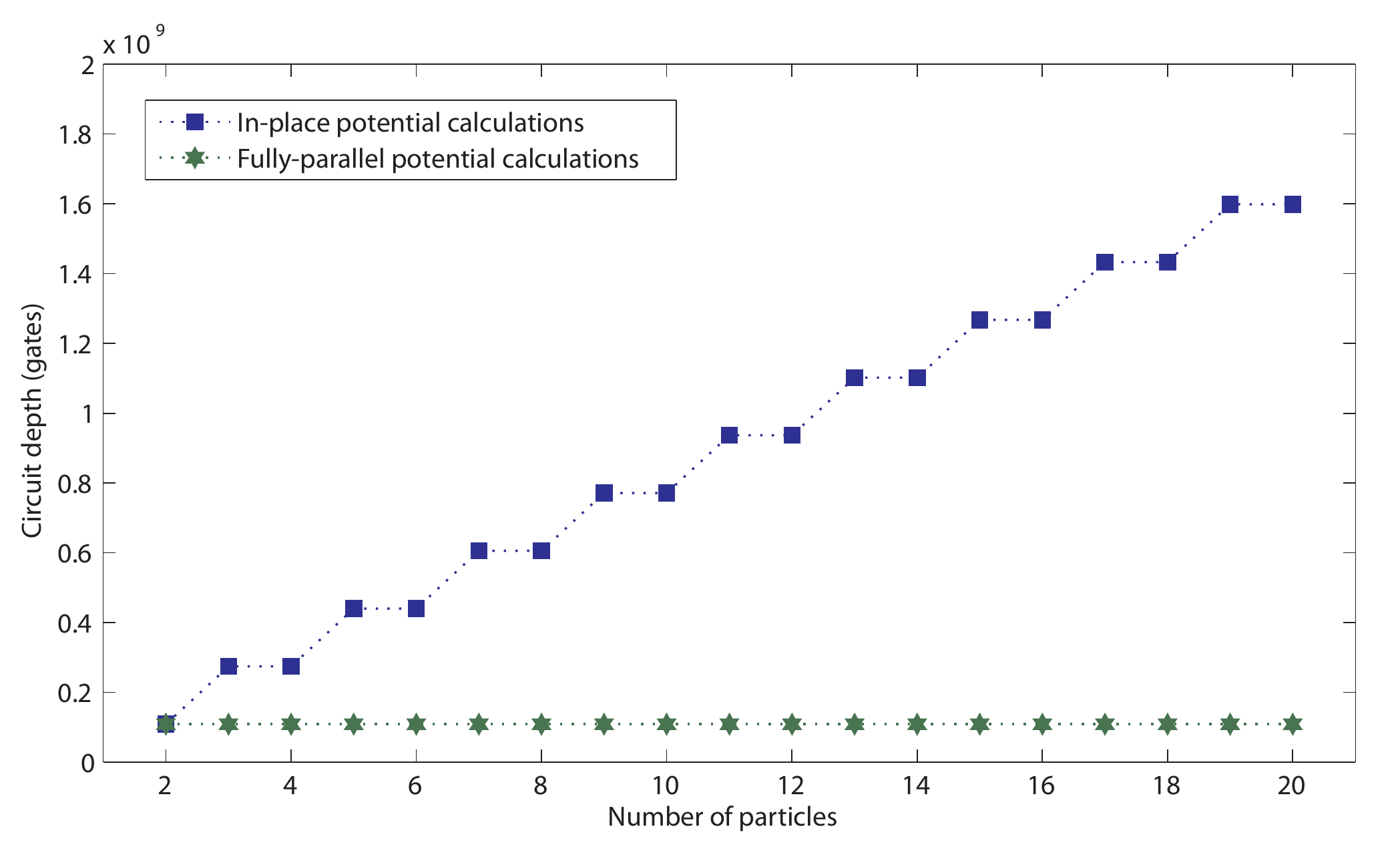}\\
  \caption{Color.  Circuit depth for two instances of first-quantized simulation.  The in-place calculation of potential energy computes each pairwise Coulomb interaction in sets of non-overlapping particle pairs, and both the depth and number of qubits required increase linearly with number of particles.  The fully-parallel calculation creates many copies of the wavefunction to permit the potential energy to be determined in constant time, at the expense of requiring substantially more application qubits (quadratic in number of particles).  In both cases, the wavefunction precision along any spatial dimension is 10 qubits, and the simulation uses 1023 time steps for 10 bits of precision, or $\sim 3$ significant figures.}
  \label{FirstQuant_depth}
\end{figure}

\subsection{Resource analysis for first-quantized molecular simulations}
The advantage of using the first-quantized approach is that the errors of the simulation are systematically improvable by increasing the spatial precision of the wavefunction and the temporal precision of the timesteps.  However, calculating kinetic and potential energy interactions requires quantum arithmetic circuits and phase rotations, which together require substantial resources in terms of fault-tolerant gates and qubits.  Figure~\ref{FirstQuant_depth} shows two versions of first-quantized simulation using the techniques for parallel calculation of potential energy from the previous section.  Although constant-depth evaluation of the Hamiltonian is possible, it requires a significantly larger quantum computer to achieve the parallel calculations, so this implementation is probably best suited to large-scale quantum computers.

Examining figure~\ref{FirstQuant_depth}, note that the circuit depth at 6 particles (\emph{e.g.} LiH) is comparable to that of the equivalent PAR-based second-quantized simulation in figure~\ref{LiH_combined} while requiring many more qubits, indicating that first-quantized simulation is more appropriate for larger molecules than LiH, since the circuit depth for first-quantized simulation is asymptotically less than second-quantized as particle number is increased~\cite{Kassal2011}.  Moreover, these calculations have assumed that the spatial precision is 10 qubits for any molecules with 2 to 20 particles.  As the size of the molecule increases, the number of qubits for each dimension of the encoded wavefunction will have to increase as the molecule itself is spatially larger.  One may also choose to increase spatial resolution to achieve a higher-precision simulation.  Each of the methods we propose for improving first-quantized simulation are summarized in Table~\ref{first_quant_summary}.

\begin{table}
  \centering
  \begin{tabular}{|>{\raggedright\arraybackslash}m{2.3cm}|>{\raggedright\arraybackslash}m{3cm}|>{\raggedright\arraybackslash}m{3cm}|>{\raggedright\arraybackslash}m{3cm}|}
    \hline
    \textbf{Method} & \textbf{Description} & \textbf{Advantages} & \textbf{Disadvantages} \\ \hline
     Quantum variable rotation (QVR) & Use phase kickback to apply a fault-tolerant phase rotation to each element in a superposition, proportional to the binary-encoded value of that element.  & Reduces complexity of first-quantized simulation.  Circuit depth is essentially the same as single-qubit phase kickback, but the QVR requires substantially fewer \texttt{T}~gates than the method in figure~\ref{QVR_bitwise}. & Not the minimal depth achievable, such as with PARs. \\ \hline
     Parallel evaluation of potential energy terms & Reduce potential operator circuit depth using parallel computation. & Shorter circuit depth than calculating all $\frac{1}{2}b(b-1)$ terms individually. & Concurrent computation requires more \texttt{T}~gates simultaneously. \\ \hline
     Teleportation circuit expansion for potential operator & Use a teleportation circuit to ``control-copy'' position-basis wavefunction in constant time. & Potential operator can be evaluated in a time which is independent of problem size.  & Circuit size in qubits increases to $O(b^2)$ from $O(b)$. \\ \hline
  \end{tabular}
  \caption{Summary of methods for efficient first-quantized chemical simulation.  The quantity $b$ is the number of particles in the chemical problem, which influences algorithm resource costs.}
  \label{first_quant_summary}
\end{table}

\section{Comparing simulation methods}
\label{comparing_methods}
The prior sections illustrate that there exist numerous ways to simulate a molecular Hamiltonian, including choices between encoded representation in a quantum computer and the way fault-tolerant rotations are prepared.  The final result one desires to know is, Which method is best?  Determining an optimal approach is subjective to the quantum computing resources available, so in this section we describe how to make such a decision.

\begin{figure}
  \centering
  \includegraphics[width=\textwidth]{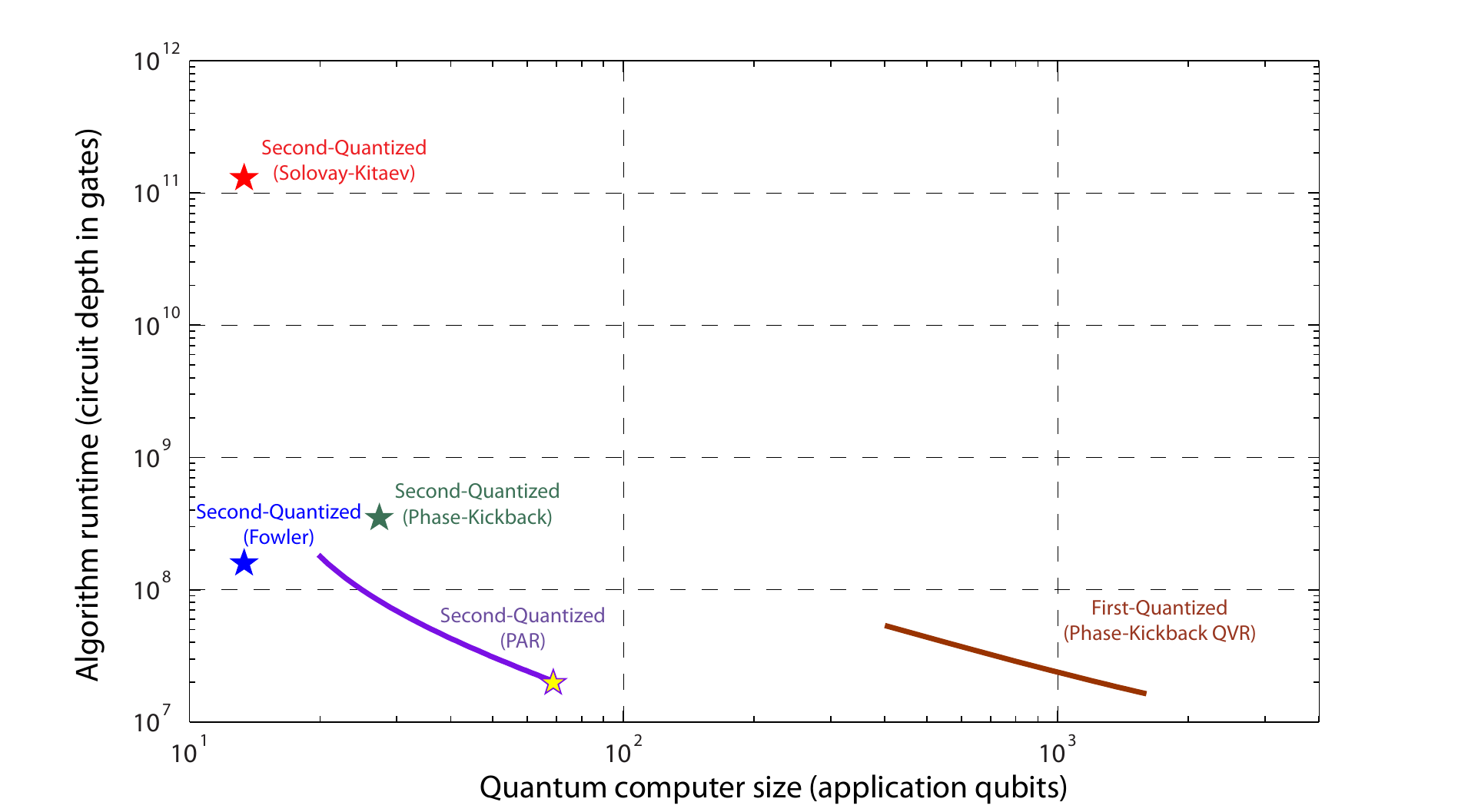}\\
  \caption{Color.  The efficient frontiers for various implementations of simulating LiH ground state energy on a quantum computer.  Each star data point corresponds to the equivalent method in figure~\ref{LiH_combined}, at rotation accuracy $\epsilon_{\mathrm{max}} \le 10^{-4}$; similarly, first-quantized simulations use QVRs with the same accuracy.  The PAR frontier (purple) and first-quantized frontier (brown) have adjustable parameters that reduce circuit depth through parallel computation at the expense of increased system size (application qubits).  For example, the PAR-based algorithm only achieves the circuit depth shown in figure~\ref{LiH_combined} when the system has 68 qubits, which is the yellow star here.}
  \label{Efficient_frontier}
\end{figure}

To visually compare different implementations of a simulation algorithm, we plot the \emph{efficient frontier} for each method in a plane defined by machine size (qubits) on the $x$-axis and execution time (circuit depth) on the $y$-axis.  The efficient frontier is the set of all points (size, depth) such that for each achievable machine size, the (achievable) depth is minimized, and vice versa.  As an example, figure~\ref{Efficient_frontier} shows the efficient frontiers of various implementations of a LiH simulation.

To determine the optimal implementation, one specifies a cost function $g(x,y)$, which associates with any point $(x,y)$ a ``cost'' to implement simulation using these parameters.  For example, cost could be the estimated engineering challenge to produce a quantum computer of size $x$ qubits combined with a penalty for the execution time of $y$ gates, which is a measure of performance.  Minimizing the cost function along each efficient frontier gives the optimal set of parameters for that particular method, and minimizing over all efficient frontiers gives the best implementation that is known to be achievable.

For the various implementations for a LiH simulation in figure~\ref{Efficient_frontier}, it seems likely that one would choose between the compact algorithm with Fowler gate sequences or the faster version with PAR sequences, which requires additional qubits to compute the necessary ancillas.  First-quantized can potentially deliver the fastest execution time here, but for this problem the number of qubits required is substantially greater.  Still, first-quantized gains an appreciable performance advantage if the number of particles is increased or if one moves to simulating time-varying dynamics~\cite{Kassal2011}.

Naturally, future algorithm advancements could produce new frontiers that are more desirable for a given cost function.  In general, one would like to make such comparisons, which can inform design decisions for quantum hardware, with full consideration of the cost to implement error correction, produce non-Clifford group gates (\emph{e.g.} \texttt{T} gates), and so forth.  However, such comprehensive system analysis is beyond the scope of this investigation; see Refs.~\cite{Isailovic08,Clark09,VanMeter09,Jones2011} for further details.

\section{Conclusions}
This paper examines the methods required to simulate chemistry on a fault-tolerant quantum computer.  A crucial operation in these algorithms is the production of phase rotations, and several approaches---phase kickback, gate approximation sequences, programmable ancilla rotations (PARs), and quantum variable rotations (QVRs)---are analyzed.  First, it should be clear that sequences generated by the Solovay-Kitaev algorithm are not nearly as efficient as the alternatives, phase kickback and Fowler sequences.  Fowler sequences are the shortest for a fault-tolerant single-qubit rotation, but the classical computing effort required to determine such sequences becomes intractable for high-precision (\emph{e.g.} $\epsilon < 10^{-6}$) rotations.  Phase kickback is a versatile technique that produces rotations comparable to Fowler's algorithm in resource usage, with the former having circuit depth $O(\log \epsilon)$ or $O(\log \log \epsilon)$ gates and requiring $O(\log \epsilon)$ \texttt{T}~gates.  Furthermore, the underlying circuit for phase kickback is an adder, which can be determined using efficient classical algorithms (unlike Fowler's algorithm), and phase kickback can be extended more readily to QVRs.  The PAR allows the quantum algorithm to achieve exceptionally low-circuit-depth rotations, at the expense of computing ancillas in advance (which is less efficient in terms of \texttt{T}~gates).  Finally, the QVR is particularly useful for first-quantized simulation. The relative merits of the methods for producing phase rotations are compared in Table~\ref{phase_rotations_summary}.

This investigation also examined two variants of the simulation algorithm, second-quantized and first-quantized, whose primary difference is the way wavefunctions are encoded and operated upon.  Generally speaking, second-quantized is a more compact representation, requiring fewer qubits, but it requires asymptotically longer execution times than first-quantized, measured in circuit depth, as the problem size increases in terms of independent particles to simulate.  Our results provide a more nuanced way to compare these methods by explicitly considering the possible ways to make the algorithms compatible with fault-tolerant quantum computing and the resulting resource costs incurred.  We have also introduced several improvements to the simulation algorithms.  In the second-quantized approach, one can neglect some of the integral terms smaller in magnitude than a cutoff threshold, implement the Jordan-Wigner transform in constant time, and use PARs to substantially reduce circuit depth, at the expense of requiring parallel production of the pre-computed PAR ancillas.  In first-quantized, we demonstrated how to produce QVRs with arbitrary scaling factor, as well as how to parallelize the calculation of the potential energy to time linear in system size (without increase in qubits) or to constant time (requiring a number of qubits that grows quadratically instead of linearly with number of particles simulated).  The methods we present for efficient chemical simulation on quantum computers are summarized in Tables~~\ref{second_quant_summary} and~\ref{first_quant_summary}.

Although we have focused on simulating quantum chemistry, these methods can be extended to simulating other Hamiltonians on quantum computers, such as spin lattice models~\cite{Lidar1997}, lattice gas automata~\cite{Boghosian1998} and lattice gauge theories~\cite{Byrnes2006}, or quantum chaos theories~\cite{Levi2003}.  Moreover, the fault-tolerant rotations could find application in other quantum algorithms, including any which require a Fourier transform.  This investigation provides a flexible set of methods for making simulation algorithms practically realizable on fault-tolerant quantum computers.

\ack
The authors would like to thank: Kevin Obenland for suggesting improvements to quantum circuits; Paul Pham for providing assistance with Solovay-Kitaev code; Aram Harrow and Isaac Chuang for helpful discussions on Solovay-Kitaev; and Austin Fowler for providing code for minimum-length approximation sequences.  This work was supported by the National Science Foundation CCF-0829694, the Univ. of Tokyo Special Coordination Funds for Promoting Science and Technology, NICT, and the Japan Society for the Promotion of Science (JSPS) through its ``Funding Program for World-Leading Innovative R\&D on Science and Technology (FIRST  Program).'' \mbox{NCJ} was supported by the National Science Foundation Graduate Fellowship.  \mbox{M-HY} and \mbox{AA-G} acknowledge support from the Air Force Office of Scientific Research, award no. FA8721-05-C-0002.  \mbox{AA-G} acknowledges support from the Alfred P. Sloan Foundation, and the Camille and Henry Dreyfus Foundation.

\appendix

\section{Transforming the phase kickback register}
\label{transform_gamma}
In some situations it is useful to change the $k$-value in $\ket{\gamma^{(k)}}$, the phase kickback ancilla register (see Eqn.~(\ref{QFT})).  Without control over $k$, the quantum variable rotation in Section~\ref{QVR_section} would require solving Eqn.~(\ref{mod_equation}) in a quantum circuit; this step would in turn require a multiplication operation, which can be expensive in terms of quantum gates.  We deviate here from Ref.~\cite{Kitaev2002} and propose a simple way to avoid the expensive operations associated with modular multiplication.  The specific ancilla state $\left|\gamma^{(1)}\right\rangle$ does not require an additional circuit to solve Eqn.~(\ref{mod_equation}), so we create this state explicitly using a simple transform $\left|\gamma^{(k)}\right\rangle \rightarrow \left|\gamma^{(1)}\right\rangle$. We begin by factoring the $\left|\gamma^{(k)}\right\rangle$ register into individual qubits (note that all such states are separable, \emph{i.e.} not entangled):
\begin{eqnarray}
\label{SK_init}
|\gamma^{(k)}\rangle  & = &  \frac{1}{\sqrt{N}}\sum_{y=0}^{N-1}e^{-2\pi i ky/N}\left|y\right\rangle \nonumber \\
                                 & = & \frac{1}{\sqrt{N}}\left(\left|0\right\rangle + e^{-2\pi i k/2}\left|1\right\rangle\right) \otimes \left(\left|0\right\rangle + e^{-2\pi i k/4}\left|1\right\rangle\right) \otimes \ldots \nonumber \\
                                 & & \otimes \left(\left|0\right\rangle + e^{-2\pi i k/2^{n}}\left|1\right\rangle\right).
\end{eqnarray}
We convert this state into $\left|\gamma^{(1)}\right\rangle$ with a series of single-qubit phase rotations using the controlled addition circuit from figure~\ref{single_qubit_PK}. Since $k$ is odd, the first bit of our ancilla register is always $\frac{1}{\sqrt{2}}(\left|0\right\rangle - \left|1\right\rangle)$.  The next bit must be rotated by the phase gate $R_{Z}(\pi(k-1))$, which is either identity or $Z$, depending on $k$.  In general, the corrective gate applied to the $m^{\mathrm{th}}$ bit is the phase rotation $R_{Z}(2\pi\frac{k-1}{2^{m-1}})$, which may be produced using the preceding $m-1$ bits of the ancilla register and phase kickback.  By iterating through all qubits in the register, we complete the transformation with circuit depth $O(n^2)$ gates or less, depending on the type of adder used in phase kickback.  This procedure can be generalized to any transformation $\left|\gamma^{(k)}\right\rangle \rightarrow \left|\gamma^{(l)}\right\rangle$ for \emph{odd} integers $1 \le k,l < 2^n$, where $n$ is the number of bits in the phase kickback register.

\section{Quantum circuits for potential and kinetic energy operators in first-quantized molecular Hamiltonians}
\label{App_first_quant}
First-quantized molecular simulation represents the simulated system wavefunction on a Cartesian grid, and the Hamiltonian is calculated with digital arithmetic acting on this coordinate space.  Similar methods were discussed in the supplementary material for Ref.~\cite{Kassal2008}, but we update this analysis for the quantum variable rotation (QVR) introduced in this work.  The potential energy operator is diagonal in position basis, and it is the sum of Coulomb interactions between electrons and nuclei in the system: $\hat{V} = \frac{1}{2} \sum_{i \ne j} \hat{V}_{ij}$, where
\begin{equation}
\hat{V}_{ij} = \frac{q_i q_j}{4 \pi \varepsilon_0} \left(\frac{1}{|{\bf r_i} - {\bf r_j}|} \right)
\label{potential_term}
\end{equation}
and $q_j$ is the charge of particle $j$.  The prefactor on the RHS of Eqn.~(\ref{potential_term}) is a constant for any given pair of particles, and we can later encode this scaling factor into the QVR.  What remains is to calculate $\frac{1}{|{\bf r_i} - {\bf r_j}|}$ over the position-encoded wavefunction.  Each position register can be decomposed in Cartesian components $\ket{{\bf r}} = \ket{x}\ket{y}\ket{z}$, so for a pair of particles we calculate
\begin{equation}
\ket{{r_{ij}}^2} = \ket{\left(x_i - x_j\right)^2 + \left(y_i - y_j\right)^2 + \left(z_i - z_j\right)^2}.
\end{equation}
The required multiplication operations can be implemented using quantum adder circuits.  Next the quantity $\ket{\frac{1}{r_{ij}}}$ is calculated using the Newton-Raphson method with the iterative equation
\begin{equation}
a_{n+1} = \frac{1}{2} a_n \left(3 - {a_n}^2 {r_{ij}}^2 \right).
\label{Newton_Raphson}
\end{equation}
With suitably chosen initial value $a_0$, Eqn.~(\ref{Newton_Raphson}) converges within 5 iterations at 32-bit arithmetic, and typically less precision is required for simulation.  The register $\ket{\frac{1}{r_{ij}}}$ is used in a QVR with scaling factor $\xi = \frac{q_i q_j \delta t}{8 {\pi}^2 \varepsilon_0 \hbar}$ from above, where $\delta t$ is the time-step of this simulated evolution and an additional factor $1/{2 \pi}$ comes from Eqn.~(\ref{quantum_variable_rotation}).  Note that each component of $\ket{\frac{1}{r_{ij}}}$ is entangled to a position-basis component of the system wavefunction, so the QVR effectively kicks back a phase to the wavefunction.  Each of the steps prior to the QVR is uncomputed, and the net effect of this sequence of operations is to implement the potential energy propagator $e^{-i \hbar^{-1} \hat{V}_{ij} \delta t}$, as in \mbox{Eqns.~(\ref{compute_V}--\ref{uncompute_V})}.

The kinetic energy operator is calculated using a similar approach as the potential energy.  The kinetic energy is the sum of individual kinetic energy operators on each particle: $\hat{T} = \sum_j \hat{T}_j$, where
\begin{equation}
\hat{T}_j = \frac{{\hat{p}_j}^2}{2 m_j} = \frac{{\hbar^2 |{\bf k}_j|^2}}{2 m_j}.
\end{equation}
The quantity $m_j$ is the mass and ${\bf k}_j = {\bf p}_j/\hbar$ is the non-relativistic wavevector corresponding to particle $j$.  By performing a quantum Fourier transform along each spatial dimension of the wavefunction, the system representation is transformed from position basis to momentum basis: $\{x,y,z\} \rightarrow \{k_x,k_y,k_z\}$.  This form permits immediate calculation of magnitude squared of the wavevector:
\begin{equation}
\ket{|{\bf k}|^2} = \ket{{k_x}^2 + {k_y}^2 + {k_z}^2}.
\end{equation}
The $\ket{|{\bf k}|^2}$ register is used in a QVR with scaling factor $\xi = \frac{\hbar \delta t}{4 \pi m_j}$.  Afterwards, the intermediate registers used in the calculation of $\ket{|{\bf k}|^2}$ are uncomputed, and the end result is the operator $e^{-i \hbar^{-1} \hat{T}_{j} \delta t}$.\\

\bibliography{FT_sim_references}
\bibliographystyle{unsrt}

\end{document}